\documentstyle[twocolumn,prd,aps,epsfig]{revtex}
\newcommand{\mum}{\mbox{${\rm \mu^{-}}$ }}
\newcommand{\mup}{\mbox{${\rm \mu^{+}}$ }}
\newcommand{\gr}{\mbox{${\rm g/cm^{2}}$}}
\begin{document}
\draft
\title{Measurement of the flux of atmospheric muons with the CAPRICE94
apparatus}

\author{M. Boezio\thanks{Also at Sezione INFN di Trieste, Trieste, Italy. 
Electronic address: mirko.boezio@trieste.infn.it}, P. Carlson, T.
Francke, and N. Weber} 
\address{ Royal Institute of Technology (KTH), S-104 05 Stockholm, Sweden}

\author{M. Suffert}
\address{ Centre des Recherches Nucl\'{e}aires, BP20,
F-67037 Strasbourg-Cedex, France}

\author{M. Hof, W. Menn, and M. Simon}
\address{ Universit\"{a}t Siegen, 57068 Siegen, Germany}

\author{S. A. Stephens\thanks{Now at Department of Physics, 
University of Tokyo,
7-3-1 Hongo, Bunkyo-ku Tokyo 113-0033, Japan}}
\address{ Tata Institute of Fundamental Research, Bombay 400 005, India}

\author{R. Bellotti, F. Cafagna, M. Circella, and C.
De Marzo}
\address{ Dipartimento di Fisica dell'Universit\`{a} and Sezione INFN
di Bari, Via Amendola 173, I-70126 Bari, Italy}

\author{N. Finetti, P. Papini, S. Piccardi, and P. Spillantini}
\address{ Dipartimento di Fisica dell'Universit\`{a} and Sezione INFN di
Firenze, Largo Enrico Fermi 2, I-50125 Firenze, Italy}

\author{M. Ricci}
\address{ Laboratori Nazionali INFN, Via Enrico Fermi 40, CP 13, I-00044
Frascati, Italy}

\author{M. Casolino, M.P. De Pascale, A. Morselli,  P. Picozza,
and R. Sparvoli}
\address{ Dipartimento di Fisica dell'Universit\`{a} and Sezione INFN di Roma,
Tor Vergata, Via della Ricerca Scientifica 1, I-00133 Roma, Italy}

\author{G. Barbiellini, P. Schiavon,  A. Vacchi, and N. Zampa}
\address{ Dipartimento di Fisica dell'Universit\`{a} and Sezione INFN di
Trieste, Via A. Valerio 2, I-34147 Trieste, Italy}

\author{C. Grimani}
\address{ Istituto di Fisica dell'Universit\`{a} di
Urbino, I-61029 Urbino, Italy}

\author{J.W. Mitchell, J.F. Ormes, and R.E. Streitmatter}
\address{ Code 661, NASA/Goddard Space Flight Center, Greenbelt, MD 20771, USA}

\author{U. Bravar, R. L. Golden\thanks{Deceased.}, and S.J. Stochaj}
\address{ Box 3-PAL, New Mexico State University, Las Cruces, NM 88003, USA}

\author{{\bf 11th April 2000}}

\author{{\bf To appear in Phys. Rev. D}}
\maketitle
\begin{abstract}
A new measurement of the momentum spectra of both positive
and negative muons as function of atmospheric depth was made by
the balloon-borne experiment \mbox{CAPRICE94}. The data were collected during 
ground runs 
in Lynn Lake on the 19--20th of July 1994 and during the balloon flight on the
8--9th of August 1994. We present results that cover the momentum 
intervals 0.3--40~GeV/$c$ for \mum and 0.3--2~GeV/$c$ for \mup,
for atmospheric depths from 3.3 to 1000~\gr, respectively. Good 
agreement is found with previous measurements for high momenta, 
while at momenta
below 1~GeV/$c$ we find latitude dependent geomagnetic effects.

These measurements are important cross-checks for the simulations 
carried out to
calculate the atmospheric neutrino fluxes and to understand the observed
atmospheric neutrino anomaly.
\end{abstract}
\pacs{96.40.Tv, 14.60.Pq, 14.60.Ef}
\section{Introduction}

Recent atmospheric neutrino observations by Super-Kamiokande
\cite{fuk98} and, with lower statistics, Soudan-2 and MACRO collaborations
\cite{amb98} have been used 
as evidence of neutrino oscillation.
The observed rate of neutrino interactions 
were compared to the rates calculated using the neutrino fluxes
derived from atmospheric cascade simulations~\cite{hobabu}. It was found
that the observed number of events induced by muon neutrinos 
is too few compared
to that from simulation. 
While several aspects of the measurements, such as the 
zenith angular dependence
of the observed atmospheric neutrinos~\cite{kan99}, strongly
point toward
the hypothesis of neutrino oscillations, the precise determination of the
allowed and excluded regions in the oscillation parameter space rely
heavily on the comparison of measurements with calculations~\cite{bat99}.

Recently Gaisser {\it et al.}~\cite{gai96} compared different
calculations of the atmospheric neutrino flux and concluded that 
the differences can be attributed to three main effects. First, 
the different parametrizations used of the particle production and
pion yield in
hadronic interactions of the primary cosmic rays with air nuclei. Second, 
the absolute energy spectra of the primary cosmic rays (protons
and helium nuclei). Third, the solar
modulation and geomagnetic effects. These differences reflect the existing
experimental uncertainties on these topics. The available accelerator data on
the pion yield in interactions of hadrons with air nuclei
are limited and the results used in the simulations do not 
cover the whole phase
space particularly at low values of the Feynman x.
Older measurements of the primary cosmic ray
spectra \cite{ryan72} differ by as much as 40\% from more recent observations
\cite{seo,boe99a,bel99} which are in
agreement at the level of 10\%--20\%, compatible with the statistical and
systematic uncertainties of the measurements. Besides the uncertainties on the
absolute values, the primary cosmic ray fluxes, particularly 
below 10~GeV/n, are affected by the periodic solar
activity and the geomagnetic field. Furthermore, the geomagnetic field affects
the distribution of the cascade in the atmosphere and these effects are not
accounted for in most of the simulations which use a 1-dimensional approach,
that is all secondaries are assumed to be collinear with the direction of the
primary  particles. Gaisser {\it et al.}~\cite{gai96} concluded that the
calculated neutrino interaction rates are uncertain at the level of $\pm 30\%$.
They also concluded that because of the cancellation of errors,
the ratio of $\nu_{\mu}$ to $\nu_{e}$ has a 
smaller uncertainty, on the order of $\pm 5\%$.

The majority of the Super-Kamiokande events are in the Sub-GeV region
\cite{kan99} where the above uncertainties are the most critical. 
Measurements of the flux of atmospheric muons provide a powerful tool for
cross-checking the cascade simulations.
Nearly all the sub-GeV neutrinos events
originate from the \( \pi \rightarrow \mu \rightarrow e \) decay chain.
Neutrinos with an energy of 1~GeV are generated, on average, by muons with an
energy of about 3~GeV. This interaction typically takes place at altitudes 
between 12 and 26 km (200 to 20~\gr). Hence, muon measurements
must be performed over a broad energy range, from a few hundred MeV to tens of
GeV, and over an extended range of atmospheric depths in order to be suitable
for use in cross-checking.

We present in this paper results on the muon spectra in the atmosphere 
obtained with the CAPRICE94 instrument from
ground (360~m) to float (36--38~km) altitude. 
First results on this analysis were reported earlier \cite{boe99b,kre99}. 
In \cite{boe99b} we compared
our measured fluxes to results of simulations and concluded that the
calculations overestimated our measured muon fluxes; the differences depending
on momentum and atmospheric depth. Theoreticians are introducing changes in the
simulations procedures, e.g. changing from one-dimensional to three dimensional
interaction models, to account for these discrepancies. Therefore,
in this paper we do not make any comparisons with
simulation results. 
In this paper we provide details of the data analysis and present 
our final results. We describe the detector system in this experiment 
in section 2, 
the data analysis in section 3 and the results and discussion comprise 
section 4.

It is worth mentioning that the primary cosmic ray hydrogen and helium
spectra \cite{boe99a} also were measured with
the CAPRICE94 instrument. 
These can be used as the input spectra for the cascade 
simulations in order to reduce the overall systematic uncertainties associated
with the comparison of observed and calculated muon fluxes.

\section{Detector system}

\begin{figure}
\mbox{\epsfig{figure=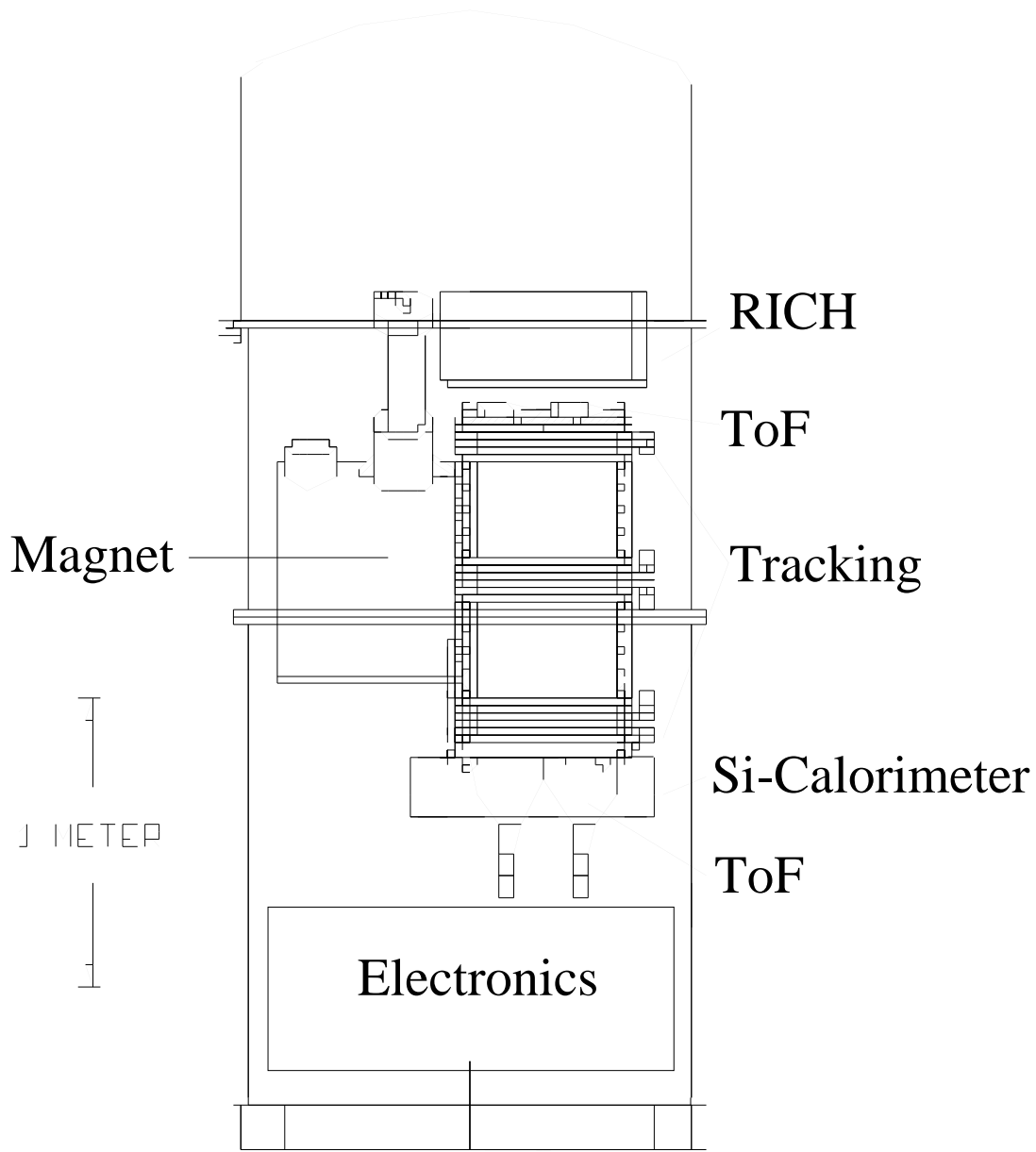,height=15cm}}
\vspace*{2.0cm}
\caption[CAP94-fig.ps]{Schematic view of the CAPRICE94 apparatus.
\label{FigGon}}
\end{figure}

   Figure~\ref{FigGon} shows the NMSU-WiZard/CAPRICE spectrometer that was
operated on ground at Lynn Lake, Manitoba, Canada (56.5$^{\circ}$ North
Latitude,
   101.0$^{\circ}$ West Longitude, 360~m altitude), in summer 1994. The payload
   was flown
   by balloon from Lynn Lake to Peace River, Alberta, Canada
   (56.15$^{\circ}$ North Latitude, 117.2$^{\circ}$ West Longitude),
   on August 8--9, 1994. The balloon reached a altitude of 
   36.0~km, corresponding to an atmospheric pressure of 4.5 mbar, in about
   three hours of ascent and 
   it floated at altitudes ranging from 38.1 to 36~km,
   i.e. residual atmosphere of 3.3--4.6~\gr, for about 23 hours.
   The apparatus included from top to bottom: a Ring Imaging Cherenkov (RICH)
   detector, a
   time-of-flight (ToF) system, a superconducting magnet spectrometer equipped
   with multiwire
   proportional chambers (MWPC) and drift
   chambers (DC) and a silicon-tungsten imaging calorimeter.

   The 51.2$\times$51.2 cm$^{2}$ RICH detector \cite{car94}, used a 1~cm thick
    solid sodium flouride (NaF) radiator with a
   threshold Lorentz factor of 1.5,
   and a photosensitive MWPC with pad readout to detect the Cherenkov light
   image and hence measure the velocity of the particles.

   The time-of-flight system had two layers one above and one below the
   tracking stack, each layer made of two 1~cm thick 25$\times$50
   cm$^{2}$ paddles of plastic scintillator.
   Each paddle had two 5~cm diameter photomultiplier tubes, one at each end.
   The distance between the two scintillator layers was 1.1~m.
   The time-of-flight system was used to
   give a trigger, to
   measure the time-of-flight
   and the ionization (dE/dX) losses of the particles.
The trigger was
a four-fold coincidence between two photomultiplier tubes
with signals in the
top paddle and two in the bottom scintillator paddle. The threshold of each
photomultiplier tube
was set high enough to eliminate noise, but low enough to
provide an efficiency of nearly 100\% to trigger
minimum ionizing particles.

   The spectrometer consisted of a superconducting magnet and a tracking device
   equipped with 
   multiwire proportional chambers
   and drift chambers \cite{gol78}.
   The magnet consisted of a single coil of 11\,161 turns of copper-clad
   NbTi wire. The outer diameter of the coil was 61~cm and the 
   operating current
   was 120~A, producing an inhomogeneous field of about 4~T at the 
   center of the
   coil. The spectrometer provided 19 position
   measurements (12 DC and 7 MWPC) in the direction of maximum 
   bending ({\em x})
   and 12 measurements (8 DC and 4 MWPC) along the perpendicular 
   direction ({\em y}). 
   Using the position information
   together with the map of the magnetic field, 
   the rigidity of the particle was determined.

   Finally, the electromagnetic calorimeter \cite{boc96}
   consisted of eight 48$\times$48 cm$^{2}$ planes of silicon
   strip (3.6~mm wide) detectors with both {\em x} and {\em y} readouts. These
   silicon planes 
   were interleaved with layers of tungsten converters, each one
   radiation length thick.
   Taking into account all the material, the calorimeter had a total
   thickness of 7.2 radiation lengths and 0.33 nuclear interaction lengths.
   The calorimeter provided topological 
   information on both the longitudinal and
   lateral profiles of the particle's interaction as well as a measure of the
   total energy deposited in the calorimeter.

\section{Data analysis}

The analysis was based on 85800 seconds of data taken at ground, 10000 seconds
during the ascent of the payload and 60520 seconds of data taken at float
altitude under an average residual atmosphere of 3.9~g/cm$^{2}$.

\subsection{Particle selection}

The CAPRICE94 instrument was well suited to measure the muon spectra and charge
ratio in the atmosphere against a background of
electrons, protons and heavier particles.
The background in the muon sample depended strongly on the atmospheric depth.
At float the dominant background for positive particles was protons, which
outnumbered the positive muons by
about a factor of 1000. The upper end of the energy interval for the \mup
measurement was determined by the 
ability of the RICH to reject proton events. At increasing atmospheric
depths, the abundance of the proton component decreased to a few
percent of the positive muon component at ground level. For negatively charged
particles at float, the electron component was the dominant background. The
electron background rapidly decreased with increasing atmospheric 
depth becoming
smaller than the muon component by 200~\gr~ and a small fraction of the muon
component at ground level~\cite{gol95,boe98}.
Because of this varying background different selection criteria were used for
$\mu^{+}$ and $\mu^{-}$ and for ground, ascent and float data to maximize the
efficiency while keeping the rejecting power for background events at an
appropriate level.

\begin{figure}
\mbox{\epsfig{figure=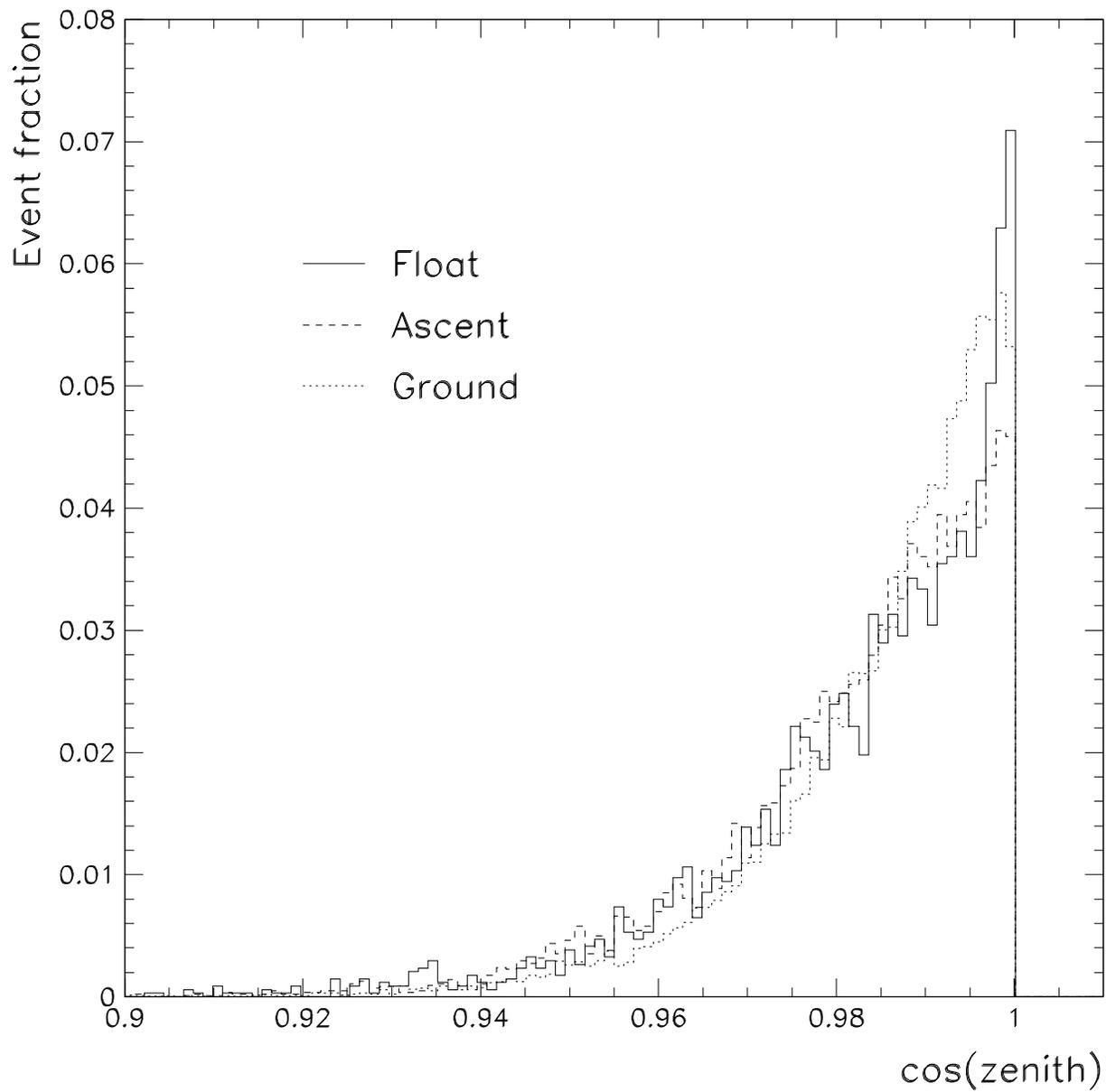,height=18cm}}
\caption[zenith.eps]{Cosine zenith angle distribution normalized to the total
number of events for muons of both signs selected 
between 0.15 and 2~GV/$c$ at ground level (dotted histogram), during the ascent
(dashed histogram) and at float (solid histogram). 
\label{zenith}}
\end{figure}

The acceptance of the instrument allowed for muons with a 
range of zenith angles
to be measured. The maximum angle was 20 degrees and the mean of the
distribution was at 9 degrees. 
Figure~\ref{zenith} shows the cosine zenith angle distribution of for muons of
both signs selected 
between 0.15 and 2~GV/$c$ at ground level, during the ascent and at float. The
distributions have been normalized for the total number of events. 
No significant change is found in the zenithal
angle distribution. It is worth pointing out that the distribution narrows as
the rigidity increases.

The selections used for identifying muons in the ground, ascent and flight data
are summarized in Table~\ref{t:selmu} and described below. The figures 
describing several of these selections show float data since at float
the background of other particles was the largest.

\subsubsection{Tracking}
   The tracking information was used to determine the 
rigidity of the particles.
In this work the
trajectory was determined by fitting only the
information from the drift chambers. This made it possible to use
the MWPC system for the efficiency estimation of the drift chambers (see
section~\ref{sss:trackeff}).
   To achieve a reliable estimation of the rigidity, a set of conditions were
   imposed on the fitted tracks:
   \begin{enumerate}
    \item At least 9 (out of 12) position measurements
      in the {\em x} direction and 5 (out of 8) in the
   {\em y} direction were used in the fit.
    \item There should be an acceptable chi-square for the
   fitted track.
    \item The estimated error on the deflection should be 
    $< 0.04$~(GV/$c$)$^{-1}$.
    \end{enumerate}
These conditions also eliminated events with more than one track in the
spectrometer.

\subsubsection{Scintillators and time-of-flight}
The ionization (dE/dX) loss in the top ToF
   scintillator was used to select minimum ionizing singly charged particles
   by requiring a
   measured dE/dX of less than 1.8 times the most probable energy loss by a
   minimum ionizing particle (mip).

   Downward moving particles were selected using the time-of-flight 
   information.
   The time-of-flight resolution of 280 ps, which was small compared to
   the flight-time of
   more than 4~ns, assured that no contamination from albedo particles remained
   in the
   selected sample. The particle's velocity ($\beta$) reconstructed from
the time-of-flight information was used
to select muons against a background of pions, protons and heavier particles.
Figure~\ref{tofmu} shows $\beta$
obtained from the time-of-flight information for positive 
particles collected at float as a
function of rigidity. Muons were selected as particles above the solid line.
The ToF $\beta$ selection was used for \mup identification for the
ascent and float portion of the data and for the \mum selection during float.

\subsubsection{The calorimeter.}
\begin{figure}
\mbox{\epsfig{figure=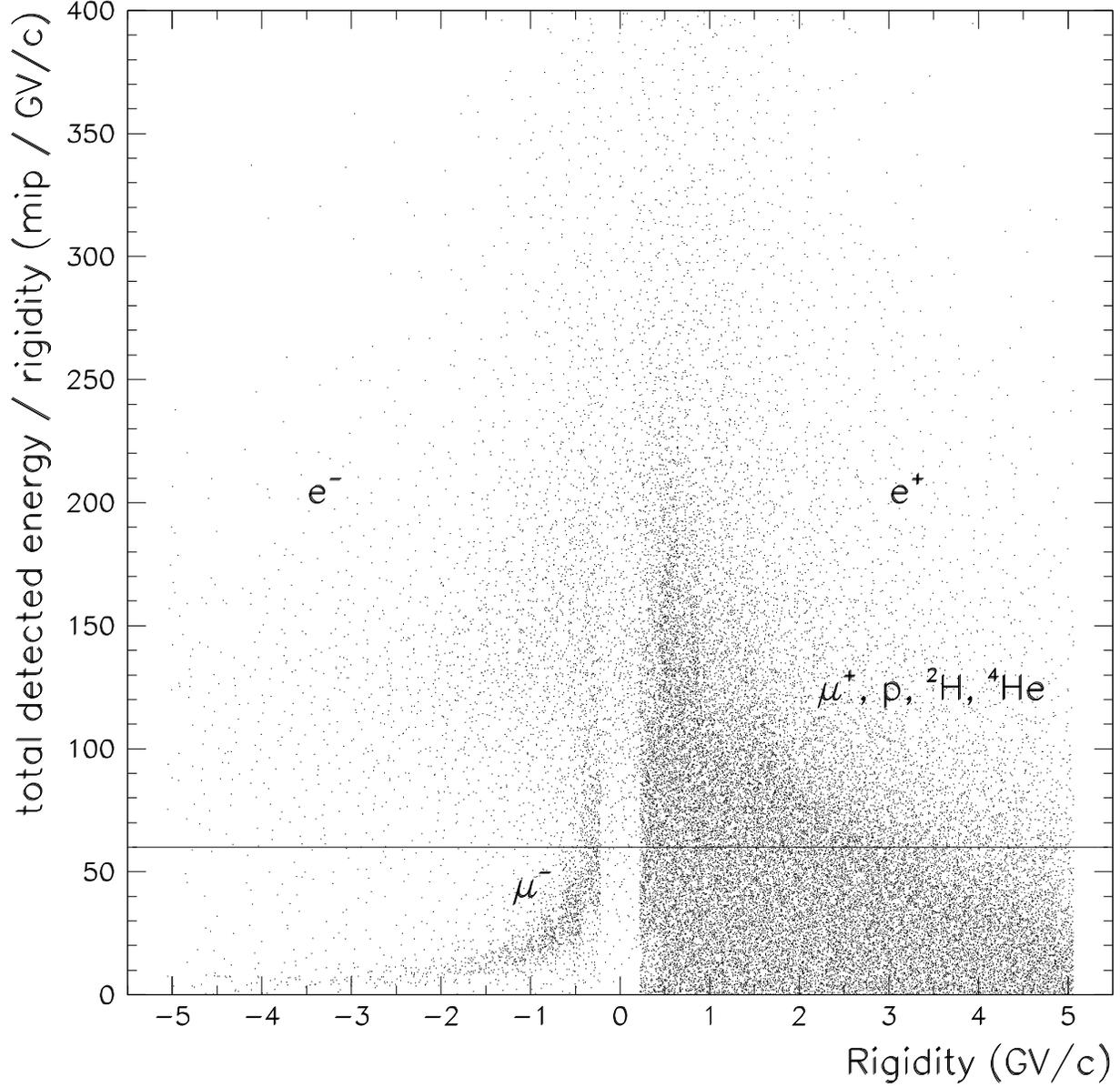,height=18cm}}
\caption[calenmom.eps]{Total detected energy loss in the calorimeter divided by
the rigidity
of all particles versus rigidity for float data. A negative sign is assigned to
the rigidity of negative particles. The two dense band are due to
non-interacting particles.
The solid line at 60 indicates the chosen upper limit used to select muons. 
The figure comprises
about 615000 events of which about 16000 are negative particles.
\label{calenmom} }
\end{figure}

With a depth of 7 radiation lengths the calorimeter could identify
non-interacting particles in a background of electrons above
a momentum of a few hundred MeV/$c$. The selection was performed by 
requiring that the topological information of the signal was consistent with
that of a single track. This was accomplished by imposing 
an upper limit to the number of
hits along the track in the calorimeter.
This selection reached its highest efficiency above 1~GeV/$c$
where electromagnetic showers were well defined in the calorimeter.
Below 1~GeV/$c$ a non-negligible electron
contamination  was present. To further reduce this background, another
selection criterion, 
based on the total detected energy in the calorimeter divided by the momentum,
was used. An upper limit
for this quantity equal to 60 mip/(GeV/$c$) was applied. Figure~\ref{calenmom}
shows this quantity as a function of rigidity for the float data.
The two dense band are due to non-interacting particles. Recall 
that muons above a few hundred MeV/$c$ are minimum ionizing particles that 
release a nearly constant amount of energy in the calorimeter.
Below 300~MeV/$c$ the calorimeter selection was
not used because of the low efficiency. At ground level 
e$^{\pm}$ amount to less than 0.1\% of the muon component above 3~GV/$c$ 
\cite{gol95,boe98} and, consequently,
the calorimeter muon selection criteria were not used above this rigidity.

The calorimeter selection criteria also rejected particles that interacted in
the calorimeter, namely pions, protons and heavier nuclei, hence contributing
to reduce their contamination in the muon sample.
However, the rejection factor for these particles was small because the
calorimeter depth was only one third of a nuclear
interaction length. Therefore, hadrons were removed using selection criteria
based on time-of-flight, scintillator and RICH information. 

\subsubsection{The RICH}
\begin{figure}
\mbox{\epsfig{figure=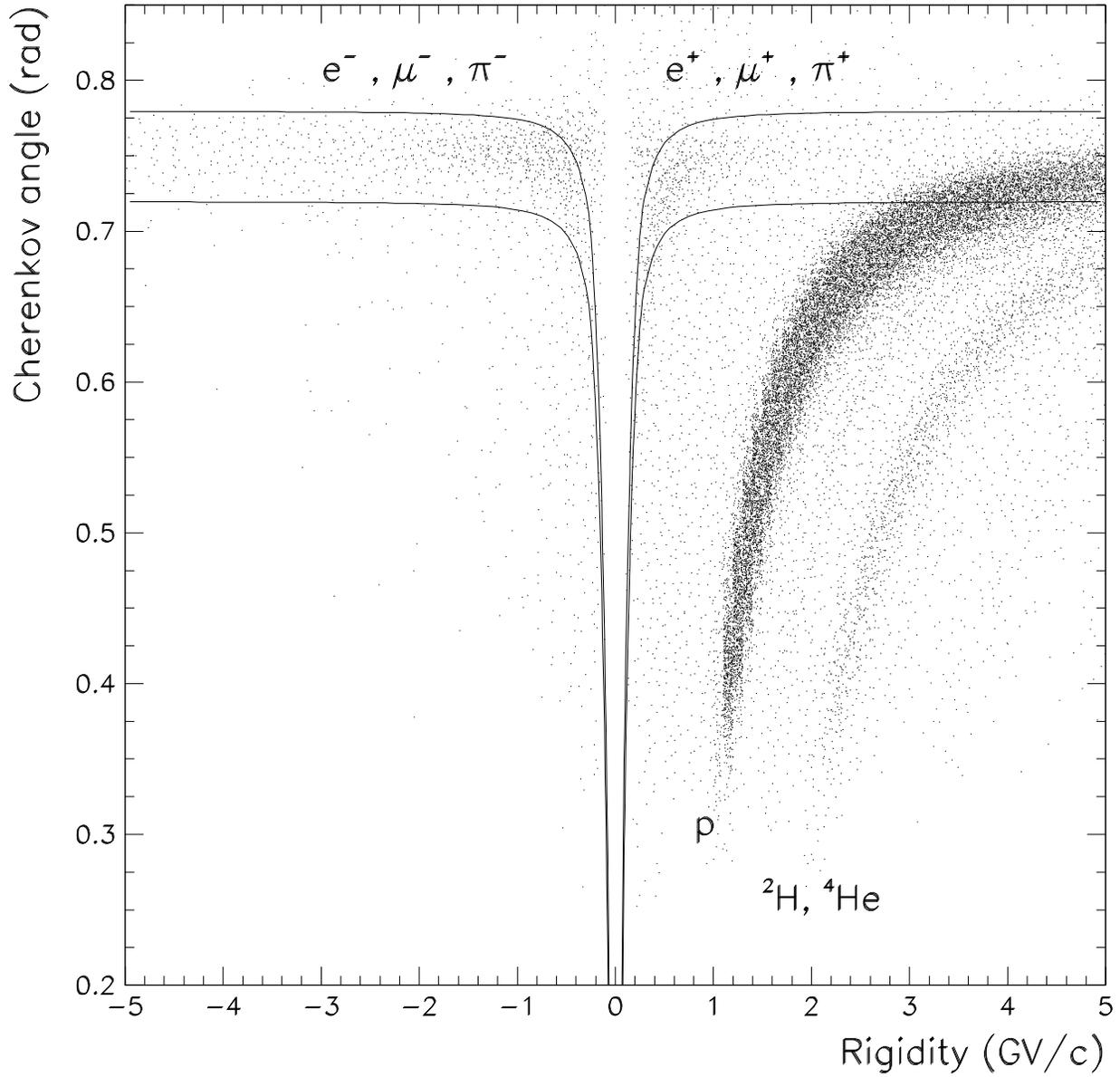,height=18cm}}
\caption[cherang.eps]{Measured RICH Cherenkov angle as a function of rigidity
for float data. A negative sign is assigned to the rigidity of negative
particles. The figure comprises about 291000 events of which about 8000 are
negative particles. The solid lines include the
events accepted as muons in the RICH selection.
\label{cherang} }
\end{figure}

The RICH was used to measure the Cherenkov angle of the particle and
thereby its velocity. The Cherenkov angle was reconstructed
from the geometrical distribution of the signals in the pad plane (for a
description of the reconstruction method see \cite{web97}).
To correctly use the RICH
information, a set of conditions
was applied on the RICH data \cite{boe98}. They were:
\begin{enumerate}
   \item Ionization from charged particles produced significantly higher signals
   than converting Cherenkov photons. To reject
   events with multiple charged particles traversing the RICH, we required 
   that an event contained only one cluster of pads with high signals.
   \item A good agreement between the particle impact position as determined
   by the RICH and the tracking system was required. The difference
   in {\em x} and {\em y} should be less than three standard deviations, which
   was rigidity dependent but typically less than 5~mm.
   \item More than 3.5 pads with signals from converted Cherenkov photons were
   required in the fit for the ground data and 
   more that 7.5 for the ascent and float data.
   \item The reconstructed
   Cherenkov angle should not deviate by more than three standard deviations
   from the expected Cherenkov angle for muons.
\end{enumerate}
Criterion~1 was introduced to eliminate events with more than one
charged particle crossing the
   RICH and this condition was applied over the entire data sets both for \mup
   and \mum. Criterion~2
eliminated events that scattered in the RICH
electronics. Criteria 3
and 4 were used to separate muons from the other particles. 
Figure~\ref{cherang}
shows the measured Cherenkov angle for flight data events selected with RICH
criteria 1, 2 and 3. The bands corresponding to the different particles are
clearly visible.
The solid lines indicate the muon selection based on the Cherenkov angle.
However, it is important to point out that the solid lines in the figure are
only indicative of the selection, since the RICH
selection was done on an event by event basis. For each event the
Cherenkov angle was obtained and the resolution of this
depended on the incident angle of the particle \cite{car94,boe98}.

These four criteria were
used for selecting muons of both signs at float and for selecting 
\mup at ground and during the ascent.
Ground and ascent \mum were selected differently: the selection was based upon
the dE/dX and calorimeter 
criteria except below 70~\gr~of residual atmosphere
where contamination of interaction products from the payload structure
(see section~\ref{ss:pionback}) was non-negligible.
For these small depths the RICH criterion~1 also 
was used. In addition the full RICH selection with criteria 1 to 4 was used
below 500~MeV/$c$ since, at low velocities, the RICH was able to separate muons
from pions and electrons.

\subsubsection{The bar}
\begin{figure}
\mbox{\epsfig{figure=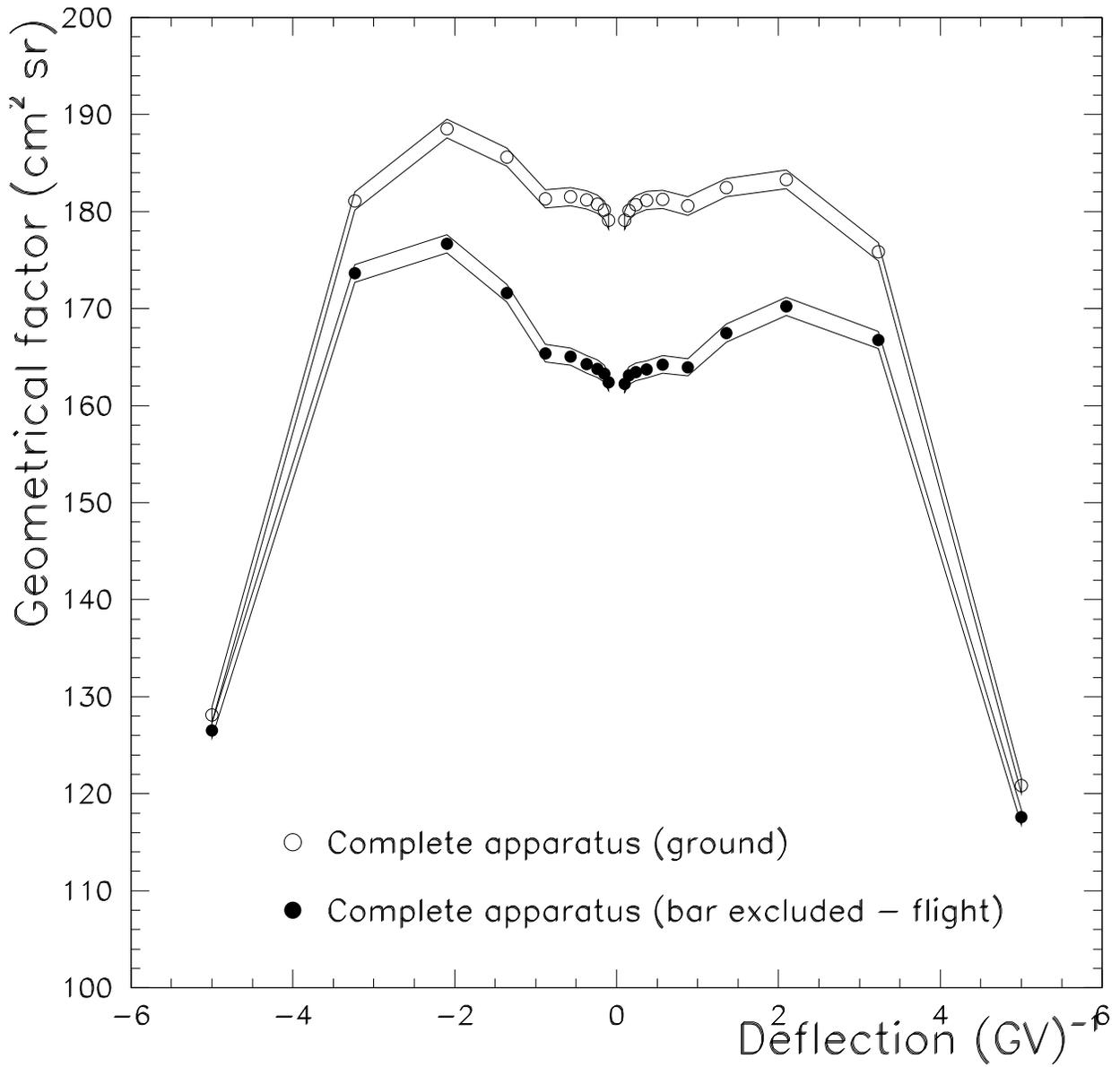,height=18cm}}
\caption{Geometrical factor as a function of deflection for ground and flight
analysis. The lines indicate one standard deviation confidence interval.
\label{geom} }
\end{figure}

   A 17 kg, 1.2 m long aluminum bar with a 7 kg steel hook in the
   centre was used to connect the payload to the balloon during the flight.
This bar was situated 2.3 m above the RICH. Hadrons 
had a non-negligible probability to interact with the material of
the bar and produce secondary particles that could be detected as muons in
the apparatus. Hence we chose to
   reject all particles with extrapolated trajectories that crossed 
the bar.
   This procedure resulted in a reduction of
the geometrical factor by about 10\% as can be seen in Figure~\ref{geom}.

\subsection{Background rejection}
The various sources of background in the muon analysis are described 
below along with the rejections criteria and surviving contamination.
This information is summarized in Table~\ref{t:back}.

\subsubsection{Electron background}
The calorimeter muon selection gave an
electron rejection factor that increased from 30 at
0.3~GeV/$c$ to more than 1000 above 1~GeV/$c$. The
RICH separated muons
from electrons with an electron rejection factor of
more than 100 at 0.1~GeV/$c$ decreasing to 10 at 0.3~GeV/$c$ and to 1 at 
0.7~GeV/$c$. 
Since the
electron flux is higher than the muon one only at high altitudes where the
electron primary component dominates and is about a factor four larger than the
muon flux \cite{boe98,boe20},
the electron and positron contaminations surviving the muon selection were
assumed negligible.

\subsubsection{Proton background.}
\label{sss:prback}
The ToF $\beta$ rejection factor for protons below 1~GeV/$c$ was greater than
4000 
decreasing to about 300 at 1.2~GeV/$c$. The RICH proton rejection factor was
greater than 2000 below 1.5~GeV/$c$, about 1000 at 2~GeV/$c$ and decreased 
to about 1 at 5~GeV/$c$.
Below 1.5~GeV/$c$, the combined effect
of the time-of-flight and RICH selection criteria reduced the proton
contamination to a negligible fraction of the selected muon sample. At higher
momenta, because of 
the strong variation of the proton flux, the
contamination was dependent on the atmospheric depth. At float altitude the
proton contamination became increasingly important above 1.5~GeV/$c$ and it 
dominated the muon sample above 2.5~GeV/$c$.
Hence, we conservatively limited the positive muon measurements to a momentum 
range between 0.15 and 2~GeV/$c$. In this range the proton contamination was
negligible at all atmospheric depths except at float.
In the float data we assumed, as a worst
case, that all the singly charged particles were protons. Applying the
rejecting power of the calorimeter, dE/dX, ToF beta and RICH to this sample
we found that a small (less than 20\%) proton
contamination survived the muon selection in the 
rigidity range 1.5 to 2~GeV/$c$.
This contamination was subtracted from the positive muon sample at float.
For ground data the range for \mup was extended to higher momenta as
presented in \cite{kre99,boe98}. 
The proton contamination was negligible at ground level below 3~GeV/$c$, while
above 3~GeV/$c$ it was calculated and subtracted from the positive muon sample.
This calculation was made by rescaling the number of interacting 
particles in the calorimeter with factors obtained from data at float. The
proton candidates were selected if they had a hadronic interaction in the
calorimeter. The contamination from muons in the interacting proton sample was
studied using negatively charged ground 
data, while the contamination from electromagnetic showers was negligible at 
momenta greater than 3~GeV/$c$ \cite{gol95,boe98}.
The efficiency of this proton selection was estimated using a
sample of singly charged particles
at float that was assumed to be composed of protons. 

In summary, during the ascent positive muons were selected
free of proton contamination up to 2~GeV/$c$. At float a small
proton contamination was subtracted from the positive muon sample between 1.5
and 2~GeV/$c$ while at ground the range was
extended to much higher momenta subtracting the small estimated proton
contamination \cite{kre99,boe98}.

\subsubsection{Heavier nuclei background.}
The case of deuteron background was very similar to the proton one.
Helium and heavier nuclei were mainly rejected by the dE/dX selection. The
remaining fraction was eliminated using the other selection methods and their
contamination was determined to be negligible.

\subsubsection{Meson background.}
\label{ss:pionback}
Because mesons (pions and kaons) produced in the atmosphere above the payload
decay rapidly, they represented a small ($< 2\%$) fraction of events compared
to the muon flux (e.g. see~\cite{bad77}). However,
pions produced in the payload could still present a non-trivial contamination.
We have undertaken a careful analysis of the local pion background in order to
quantify their abundance.

\begin{figure}
\mbox{\epsfig{figure=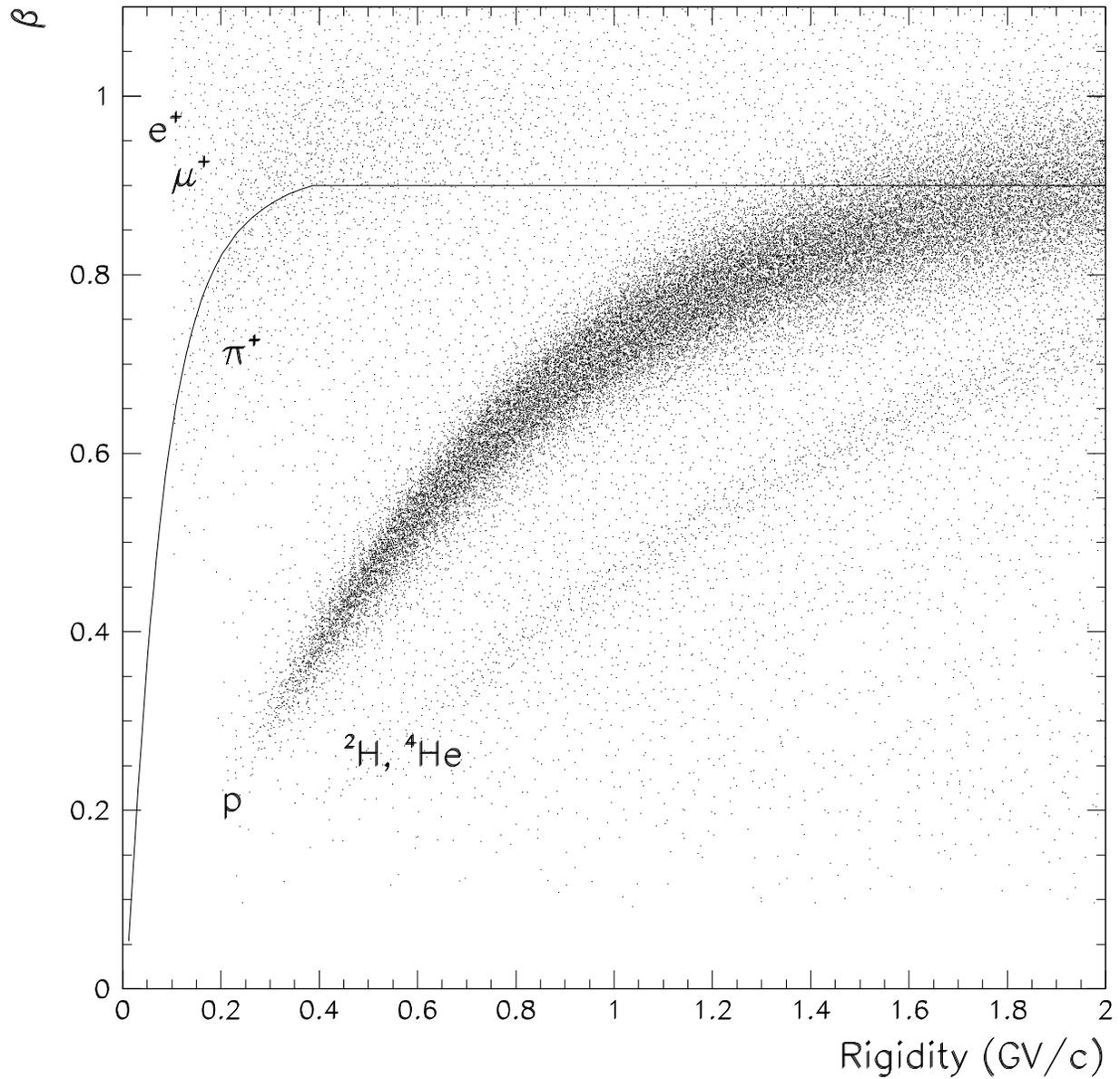,height=18cm}}
\caption[tofmu.eps]{The distribution of $\beta$ from the time-of-flight
information as a 
function of rigidity for float data. The solid line is the lower limit of
the muon selection. The distributions corresponding to the various particles
are labelled according to the particle's species.
The figure comprises
about 379000 events of which about 347000 are protons, 9000 deuterons and 6000
helium nuclei.
\label{tofmu} }
\end{figure}

Evidence for a pion contamination is visible at very low rigidities in
Figure~\ref{tofmu}. These pions,
because of their low energies, were presumably produced locally 
in the RICH or in the dome (part of the gondola above the RICH).
However, it is
important to stress that no RICH conditions were applied to select the
events in Figure~\ref{tofmu}: the RICH selection rejected
multiply charged tracks, particles interacting in the RICH as well as 
pions below 500~MV/$c$. 
Also the dE/dX selection rejected multiply charged tracks
and this leaves the dome as the source of pions
produced in the interaction with high energy cosmic rays. 
The interactions occurred at such an angle that the high energy
secondaries missed the active volume of the instrument, but a low energy pion 
was produced at a large angle and passed through the instrument 
appearing like a singly charged particle rather than part of a shower. 
In studying the pions below 
$< 0.2$~GV/$c$, where they can be identified using the time-of-flight, we found
no evidence of a preferred incoming direction.
This is consistent with our interpretation of these events since 
the distribution of 
material in the gondola was symmetric
with respect to the azimuthal angle. 
The locally produced pion flux entering the
apparatus should decrease quickly with energy
because of the emission towards the forward direction and the lower probability
of having only one charged particle with 
high momentum traversing the detectors.

To test the correctness of this conclusion the
following approach was adopted.
Events were selected from float data with:
\begin{enumerate}
    \item multiple pad ionization clusters in the RICH;
    \item multiply charged signal in the top scintillator;
    \item rigidity (R) interval: $0.5 < R < 2$~GV/$c$;
    \item Cherenkov angle of a $\beta \approx 1$ particle;
    \item time-of-flight of a $\beta \approx 1$ particle;
    \item dE/dX signal of a minimum ionizing particle in the bottom
    scintillator.
\end{enumerate}
With these criteria
$\beta \approx 1$ particles belonging to a shower
initiated closely to the top of the apparatus were selected,
hence with high probability
secondary pions, which entered the calorimeter as singly charged. The
criteria 3 to 6 ensured that the selected events
were indeed $\beta \approx 1$ particles.
This sample of locally produced pions was used to estimate
their interaction
probability in the calorimeter (or, more precisely, 
the probability of selecting
interacting pions in the calorimeter). Then, the same hadronic interaction
selection for the calorimeter was applied to the muon sample selected
without using the calorimeter muon criteria.
Two samples were obtained, one for positive and one for negative pions.
These events (about 100 in total) were visually scanned with a
graphic program. In this way
misidentified muons and, especially, electrons were rejected from the samples.
Then,
the remaining pion numbers were rescaled by the calorimeter 
selection efficiency
thus obtaining the number of pions in the muon samples. The result was that at
float altitude 
pions could account for a maximum of 20\% between 0.5 and 1~GeV/$c$ 
and for less
than 10\% above 1~GeV/$c$ of the muon flux, irrespective of the sign of the
charge. We give
this result as an upper limit because the procedure is likely to
overestimate the number of pions (some of the selected pions could be muons,
etc.). For this reason, it was not subtracted from the muon flux and it should 
be considered as a systematic uncertainty of the measurement. It is
important to stress that this uncertainty, due to the similar pion 
contamination in the \mup and \mum samples, affects the \mup to \mum
ratio less than the corresponding fluxes.

\subsubsection{\bf Conclusion on background}
Clean \mup and \mum samples were selected from 0.15 to 0.4 GeV/$c$. Above this
momentum a non-negligible contamination of locally produced pions could be
present. For the float data, this contamination was less than
20\% of the muon flux between 0.5 and 1~GeV/$c$ decreasing to less than
10\% above 1~GeV/$c$. For larger
atmospheric depths the locally produced pion flux decreased quickly 
due to the decrease of the interacting proton and helium nuclei, specially at
large zenith angle, 
while the muon flux increased with increasing depth 
at least up to 100~\gr. Hence the
locally produced pion contamination was assumed
negligible at all depths except at float.

At float a small proton contamination was subtracted from the positive muon
sample between 1.5 and 2~GeV/$c$. At ground the momentum range was extended to
much higher value with a subtraction of a small proton contamination.

\subsection{Efficiency determination}
\label{ss:efficiency}
In order to accurately determine the
fluxes of the various types of particles, the
efficiency of each detector was carefully studied using both ground and flight
data.
To determine the efficiency of a given detector, a data set of muons was
selected by the remaining detectors. The number of muons correctly 
identified by the detector under test divided by the number of
events in the data set provided a measure of the efficiency. This procedure was
repeated for each detector.
The efficiency of each detector was determined as a function of rigidity in a
number of discrete bins and, then, parameterized
to allow an interpolation between bins.
This parameterization introduced a systematic error on the efficiency of each
detector. Since the parameters were correlated, the error on the efficiency
was obtained using the error matrix of the fit for each detector when
correcting the measured flux for the detector efficiencies.

\subsubsection{Tracking efficiency}
\label{sss:trackeff}
\begin{figure}
\mbox{\epsfig{figure=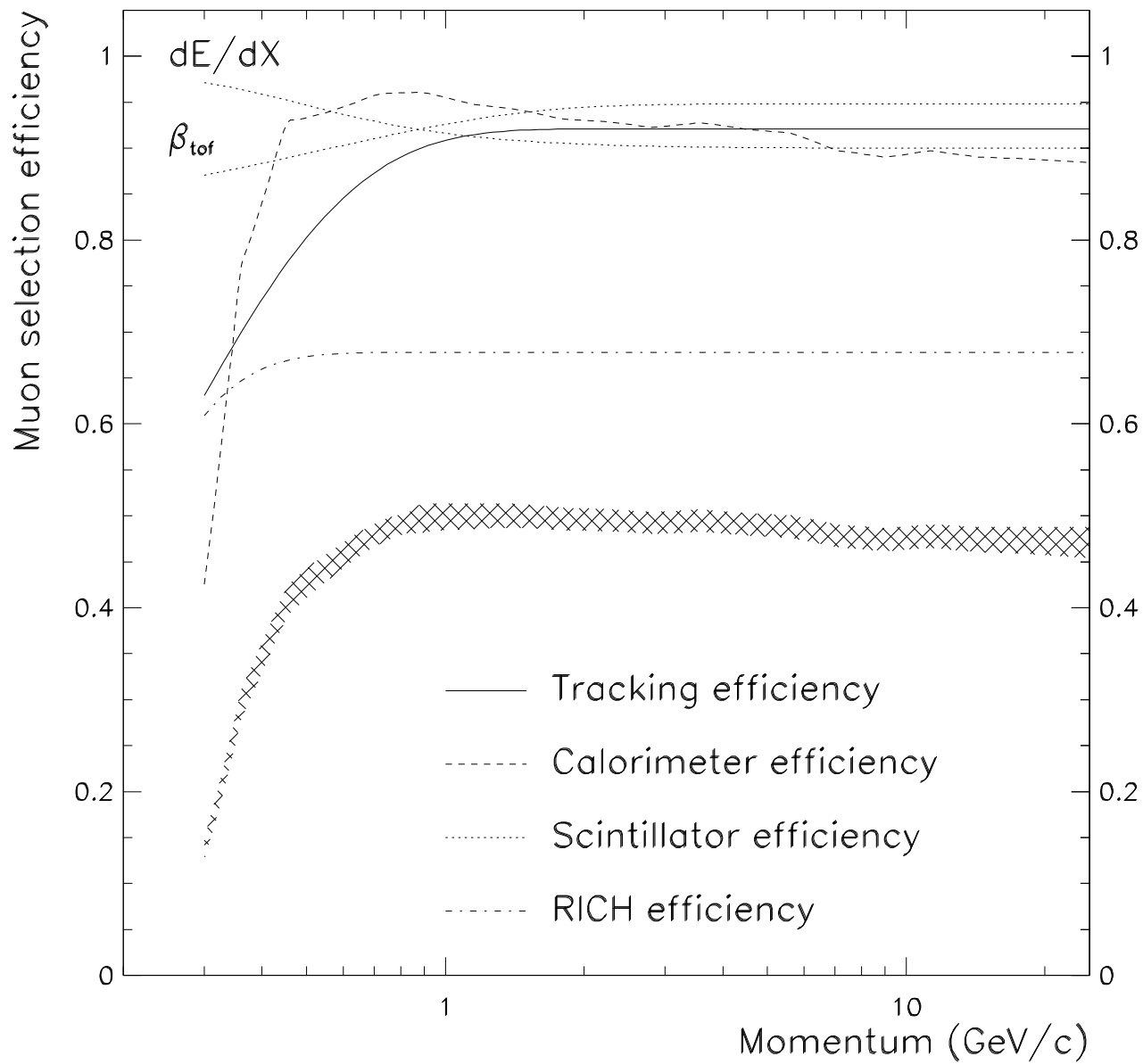,height=18cm}}
\caption{Muon selection efficiencies for float data as a function of rigidity.
The hatched area indicates one standard deviation confidence interval of the
combined efficiency.
\label{effmu} }
\end{figure}

The drift chamber tracking efficiency was obtained using negative
singly charged particles selected by
the other detectors similarly as done in \cite{boe20}.
A sample of singly charged particles was selected by requiring a single
ionization cluster in the RICH and a dE/dX signal in the top
scintillator typical of a minimum ionizing particle.
From this sample, negatively charged events were selected by
requiring a negative deflection from the fit to the MWPC
trajectory measurements.
The contamination of spillover
protons was eliminated by requiring that
the measured impact positions in the RICH and calorimeter
agreed with the positions as obtained by extrapolating the particle
trajectory derived from the MWPC fit.
The resulting sample of negative singly charged particles was used
to determine the efficiency of fitting
tracks in the drift chamber system. The solid line in Figure~\ref{effmu} shows
the tracking efficiency at float altitude.
The efficiency varied from ground to float altitude. At ground it was
$\simeq 95\%$ above 1~GeV/$c$, then just after the launch it was $\simeq 87\%$
increasing to $\simeq 93\%$ at float altitude.

Biases in the efficiency sample were studied using protons (see~\cite{web97}).
It was found that the criteria used for fitting the tracks
using the MWPC slightly reduced the number of scattered tracks in the sample.
In order to account for this reduction a systematic uncertainty of 2\% was
introduced \cite{boe98}.

Possible charge sign dependence of the efficiency was studied using both
the flight data and the data taken on the ground
before the flight. No significant dependence was found above 0.3~GV/$c$
\cite{boe98}.

\subsubsection{Scintillator efficiency}
The dE/dX and $\beta$ selections were studied using negative events with a
minimum ionizing pattern in the calorimeter. A clean RICH signal also was 
required to reject interaction products from the sample. The dotted lines in
Figure~\ref{effmu} show the scintillator efficiency in its two selections. 
The efficiency was studied
using ground, ascent and float data and no variation was found.

\subsubsection{Calorimeter efficiency}
The calorimeter selection efficiency, shown as a dashed line in
Figure~\ref{effmu}, was obtained using ground data. The result
was cross checked with a simulation of the calorimeter
\cite{boe98} and an excellent agreement was found. The calorimeter efficiency
also was studied with flight data. The calorimeter only is able to separate
muons from electrons above 0.5~GeV/$c$, hence the presence of a larger electron
contamination had to be taken into account.
Inside the errors a good
agreement was found as expected since the calorimeter performances were stable
over a period of months.

\subsubsection{RICH efficiency}
On the ground and in the first part of the ascent the
RICH efficiency was obtained by selecting negative
singly charged particles, which did not interact in the calorimeter.
At float altitude a large background of interaction products did not permit us
to select an unbiased clean sample of muons, hence the efficiency obtained from
the ground data was used. This procedure was validated by comparing the RICH
efficiency for selecting electrons at ground and at float since
an unbiased clean sample of electrons could be selected using the calorimeter.
The RICH electron selection criteria 
is similar to that for muons, namely, same requirements on the impact position,
number of 
effective pads and Cherenkov angle but using the theoretical electron Cherenkov
angle. It was found that the RICH
electron efficiency for float
data reproduced the electron efficiency of ground electrons inside an
uncertainty of about 5\%. Therefore, it was reasonable to make use of the muon
RICH efficiency as obtained from the ground data 
(dashed-dotted line in Figure~\ref{effmu}) for the flight data.

\subsection{Geometrical factor}
\label{ss:geom}

The geometrical factor, determined with simulation
techniques \cite{sul71}, is shown in Figure~\ref{geom} for ground and flight
data. The difference at low deflections (high rigidities) for the two sets of
data is due to the additional geometrical constrain imposed due to the bar.

The geometrical factor was cross checked with two other methods.
One adopted the same approach as presented in~\cite{sul71} using, however, a
different
method to trace the particles: the track fitting algorithm used in the
analysis also was used
to trace the particle through the spectrometer. This method
gave the same results within 1\%, at all
rigidities. The second used a numerical integration calculation of the
geometrical factor that agreed with
the previous results within 2\% above 0.5~GV/$c$ and within 5\% below 
0.5~GV/$c$.

\subsection{Systematic uncertainties}
\label{ss:systematics}

Systematic uncertainties originating from the determination of the detector
efficiencies were included in the tables and data points as discussed in
session~\ref{ss:efficiency}.
Another possible systematic error was related to the efficiency of the trigger
system.
The fraction of each
trigger combination was compared with the simulated fraction taking into
account the position of each paddle and the magnetic field.
The excellent agreement
between the simulated and experimental fractions
permitted us to conclude that a possible systematic error due to a geometrical
inefficiency of the trigger was less than 1\%.

The residual atmosphere
above the gondola was measured by two pressure sensors owned and calibrated by
the National Scientific Balloon Facility.
The two measurements did not
coincide: their difference increased with altitude from less than 1\% to about
10\% at float. We interpreted this difference as the systematic uncertainty on
the 
atmospheric depth. This uncertainty does not affect the measurement but has to
be taken into account when comparing the measured spectra with the simulated
ones. 

From the discussion in section~\ref{ss:geom} we conclude that
the systematic error due to the geometrical factor calculation was
less than 5\% between
0.3 and 0.5~GV/$c$ and less than 2\% for rigidities higher than 0.5~GV/$c$.
For the geometrical factor calculations it was assumed that there was no
variation of the muon
intensity over the acceptance angle. The effect on the geometrical 
factor due to
the intensity variation was examined using the measured muon spectra
\cite{tsu99} and the observed zenithal distribution in our apparatus
in the rigidity range 0.2 to 1.5~GV/$c$ at ground. We found that the calculated
geometrical factor would be reduced by about 3\%.

We decided to assign a systematic uncertainty of 5\% to the RICH muon 
efficiency
at float to account for possible variations in the RICH performance
between ground and float.

The tracking muon selection efficiency varied with time during the ascent. We
determined the efficiency for seven time bins and we found that they could be
grouped in three in which the efficiency could be assumed constant. Since the
efficiency above 1~GV/$c$ varied from about 87\% at launch to 93\% at float we
believe that the systematic uncertainty of this procedure is less than 6\%.

Assuming that the systematic uncertainties discussed above are uncorrelated, we
estimated an overall systematic uncertainty,
which is momentum dependent, for ground muons decreasing from
$\simeq$~6\% at 0.3~GeV/$c$ to about 2\% above 1~GeV/$c$. 
It is worth pointing out
that the ground muon fluxes measured by the CAPRICE94 apparatus agree at the
level of 3\% with the measurements from the CAPRICE97 experiment \cite{kre99},
which was equipped with
the same superconducting magnet and calorimeter but with a different tracking
system and with a gas RICH.
For ascent and float muons
this systematic uncertainty decreased from
$\simeq$~9\% at 0.3~GeV/$c$ to about 7\% above 1~GeV/$c$.
These uncertainties were not included in the data presented in the tables and
the figures.

\section{Results}
We selected 37864~\mum and 47043~\mup between 0.2 and 120~GeV/$c$ at ground
(1000~\gr);
5081~\mum between 0.3 and 40~GeV/$c$ and 2715~\mup between 0.3 
and 2~GeV/$c$ during
the ascent (7--850~\gr); 1601~\mum between 0.18 and 20~GeV/$c$ and 2063~\mup
between 0.18 and 2~GeV/$c$ at float altitude 
(3.3--4.6~\gr, mean atmospheric depth
of 3.9~\gr). From these we obtained the
muon fluxes ($J_{\mu^{-},\mu^{+}}$) according to:
\begin{eqnarray}
J_{\mu^{-},~\mu^{+}}(P,X) & = &
\frac{1}{T_{live}(X) \times
G_{\mu^{-},~\mu^{+}} \times \epsilon(P,X) \times
\Delta P} \nonumber \\
& & \times N_{\mu^{-},~\mu^{+}}(P,X) ,
\end{eqnarray}
where $X$ is the atmospheric depth, $T_{live}$ is the live time,
$G_{\mu^{-},~\mu^{+}}$
are the geometrical factors
for \mum and \mup, $\epsilon$ is
the combined selection efficiency, $\Delta P$ is
the width of the momentum bin corrected for ionization
losses to the top of the payload, $P$ the momentum
and $N_{\mu^{-},~\mu^{+}}(P)$ is the selected
number of \mum and \mup. The fractional live time decreased from
$0.97240 \pm 0.00001$ at ground to $0.2690 \pm 0.0006$ at float altitude as
indicated in Table~\ref{t:fluxmuon}.

\begin{figure}
\mbox{\epsfig{figure=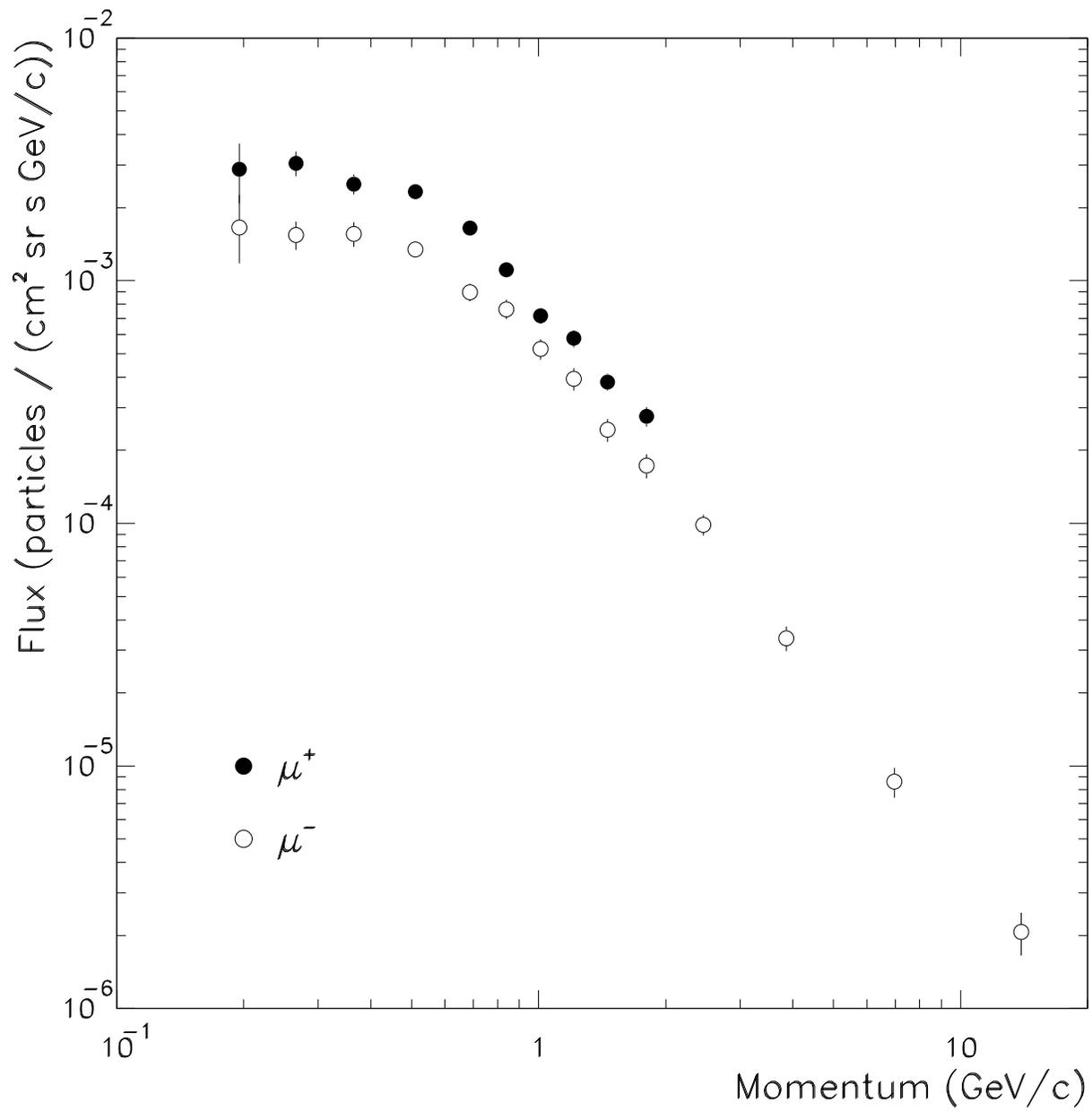,height=18cm}}
\caption{The \mup (full circles) and \mum (open circles) fluxes at float
(3.9~\gr~of mean residual atmosphere) as a function of momentum.
\label{fluxfloat} }
\end{figure}

Figure~\ref{fluxfloat} and Table~\ref{t:fluxfloat}
show the muon spectra at float, corresponding to 3.9~\gr~of residual 
atmosphere.
These muon fluxes are interesting since these muons are the
products of the first interaction between the primary cosmic rays and the air
nuclei. Hence, along with the simultaneous measurement of the primary 
spectra of
proton and helium nuclei \cite{boe99a} these data provide a useful testbench
for studying the pion production in nucleon-air interaction used in the
calculation of atmospheric showers. As discussed in section~\ref{sss:prback}
we limit the positive muon data to momenta below 2~GeV/$c$. The average muon
charge ratio on this momentum interval is $1.59 \pm 0.06$.

In Table~\ref{t:fluxmuon} we present the measured muon fluxes at several
atmospheric depths and momenta interval. The symbol FAD stands for
Flux-weighted Average Depth \cite{bel96} obtained according to:
\begin{equation}
{\rm FAD}(P) \: = \: \frac{\int \! X(t) \epsilon_{live}(t) J(P) dt}
{\int \! \epsilon_{live}(t) J(P) dt} ,
\label{eq:gram}
\end{equation}
where $\epsilon_{live}$ is the fractional live time. The depth and momentum
intervals were
chosen to match the published data of \mum flux growth curves by the MASS89 and
MASS91 experiment~\cite{bel99,bel96}.

\begin{figure}
\mbox{\epsfig{figure=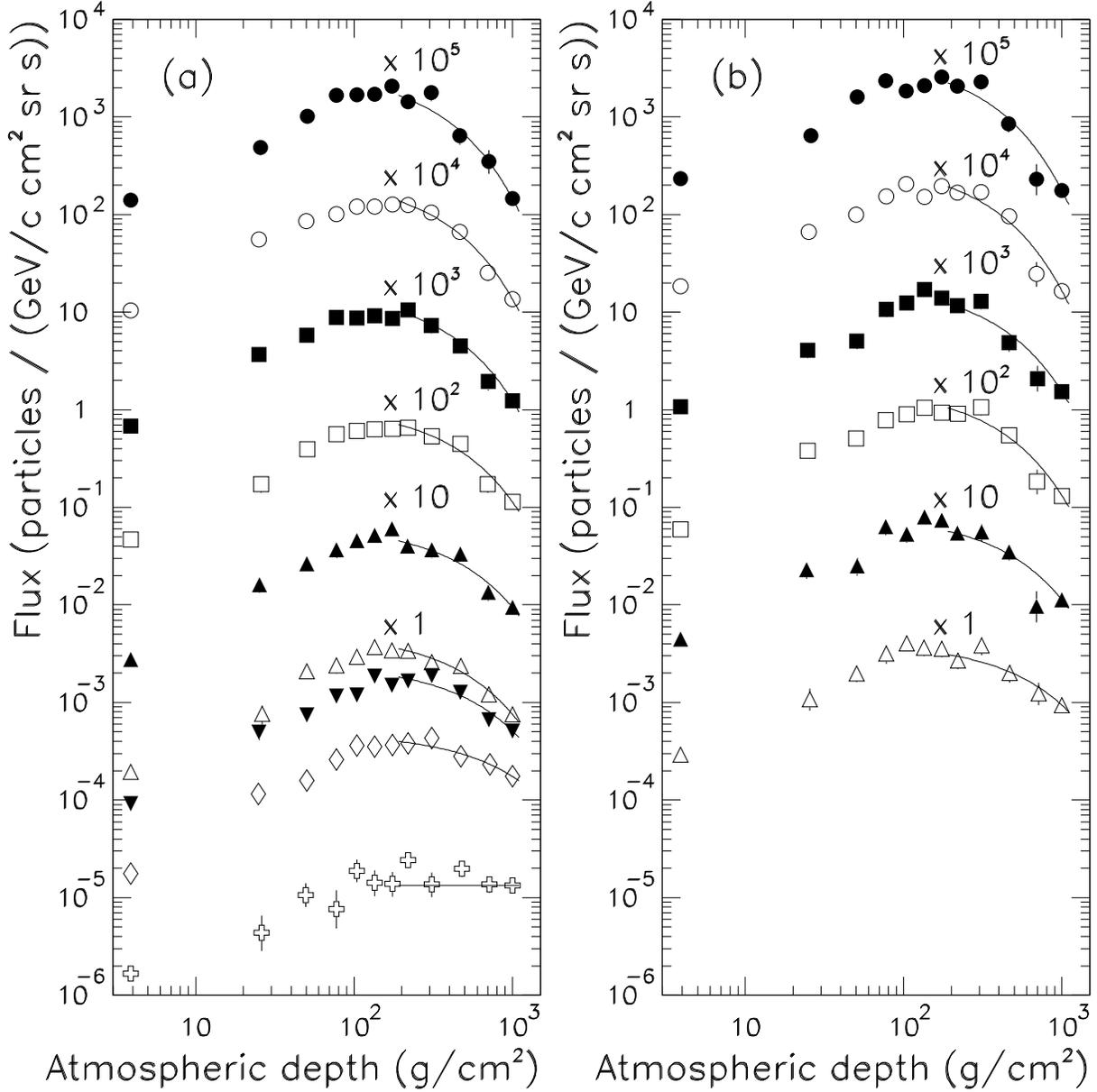,height=18cm}}
\caption{Atmospheric growth curves for a) $\mu^{-}$ in the left panel and b)
$\mu^{+}$ in the right panel. 
From top to bottom are the
momentum ranges in~GeV/$c$: 0.3--0.53 (scaled by $10^{5}$),
0.53--0.75 ($10^{4}$), 0.75--0.97 ($10^{3}$), 0.97--1.23 ($10^{2}$),
1.23--1.55 (10), 1.55--2 (1), 2--3.2 (1), 3.2--8 (1) and
8--40 (1). The $\mu^{+}$ results are shown up to 2~GeV/$c$.
The solid lines are exponential fits according to equation~\ref{eq:expo}.
\label{growthmu} }
\end{figure}

Figure~\ref{growthmu} shows the flux growth curves for (a) negative and (b)
positive muons for several momentum bins. For each momentum interval we
fitted
the data at large atmospheric depths ($X > 190$~g/cm$^{2}$) with an exponential
function \cite{bel96}:
\begin{equation}
J(P,X) \: = \: k(P) e^{-X/\Lambda(P)} ,
\label{eq:expo}
\end{equation}
where $k$ and $\Lambda$ are obtained from the fits. The resulting best fits are
shown in Figure~\ref{growthmu} as solid lines. As found in the 
MASS89 and MASS91 experiments \cite{bel99,bel96,cir97} a nearly
linear relation exists between the attenuation length ($\Lambda$) and the mean
momentum ($P$) in unit of GeV/$c$ in the 190 to 1000~\gr~ range. The relation
resulting from CAPRICE94 measurements of \mum is:
\begin{equation}
    \Lambda [{\rm g/cm^{2}}] \: = \: (263 \pm 14) \, + \, (150 \pm 15)
    \times P ,
\label{eq:lambdac}
\end{equation}
This relation holds also for the CAPRICE94 \mup flux growth
curves. It is worth pointing out that
in determining
relation~\ref{eq:lambdac} we used also the \mum fluxes at ground.
Equation~\ref{eq:lambdac} 
can be compared with the one determined from MASS89 \mum
data \cite{cir97} as:
\begin{equation}
    \Lambda [{\rm g/cm^{2}}] \: = \: (283 \pm 24) \, + \, (93 \pm 16)
    \times P .
\label{eq:lambdam}
\end{equation}
Both the above expressions reproduce the data within errors over the momentum
range of these experiments, but will differ when extended to much larger
momenta.

\begin{figure}
\mbox{\epsfig{figure=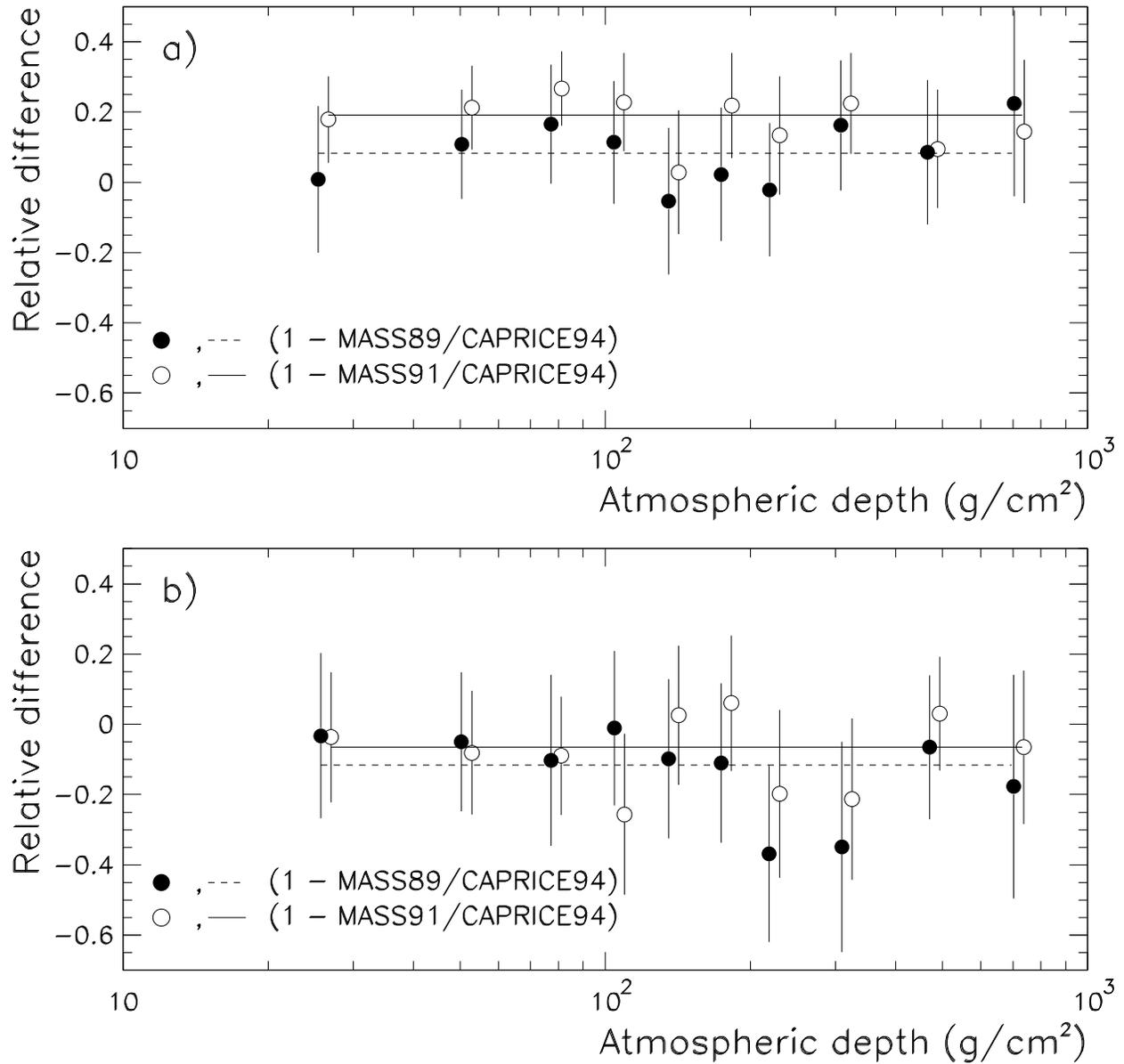,height=18cm}}
\caption{Relative difference between the \mum fluxes obtained in this analysis
and the
MASS89~\protect\cite{bel96} and MASS91~\protect\cite{bel99} experiments as a
function of atmospheric depth. The comparison is done for muon momenta
below 1~GeV/$c$ in the upper panel (a) and between 1 and 2~GeV/$c$ in the lower
panel (b).
The average difference between CAPRICE94 and MASS89 is shown as a dashed line
and between CAPRICE94 and MASS91 as a solid line.
\label{compflux} }
\end{figure}

Figure~\ref{compflux} shows the relative difference between the
\mum fluxes obtained in this analysis and the
MASS89 \cite{bel96} and MASS91 \cite{bel99} experiments as a function of
atmospheric depth. The comparison is done for muon momenta
below 1~GeV/$c$ (a) and between 1 and 2~GeV/$c$ (b). 
The dashed lines indicate the average difference between this analysis and
MASS89 and the solid lines the average difference with MASS91. 
Considering the errors in the data points, a good agreement 
is found in the 1 to
2~GeV/$c$ interval among different measurements. However, below 1~GeV/$c$ the
CAPRICE94 results are significantly higher than the results from the MASS91 
and,
to a lesser extent, from the MASS89 experiments. These differences could be
caused by solar 
activity or geomagnetic effects. In fact, the MASS89
experiment was launched from Prince Albert, Saschatcewan, Canada, during a
period of maximum solar activity while MASS91
was launched from Fort Sumner, New Mexico, and flew at an average geomagnetic
cutoff of about 4.5~GV/$c$.

\begin{figure}
\mbox{\epsfig{figure=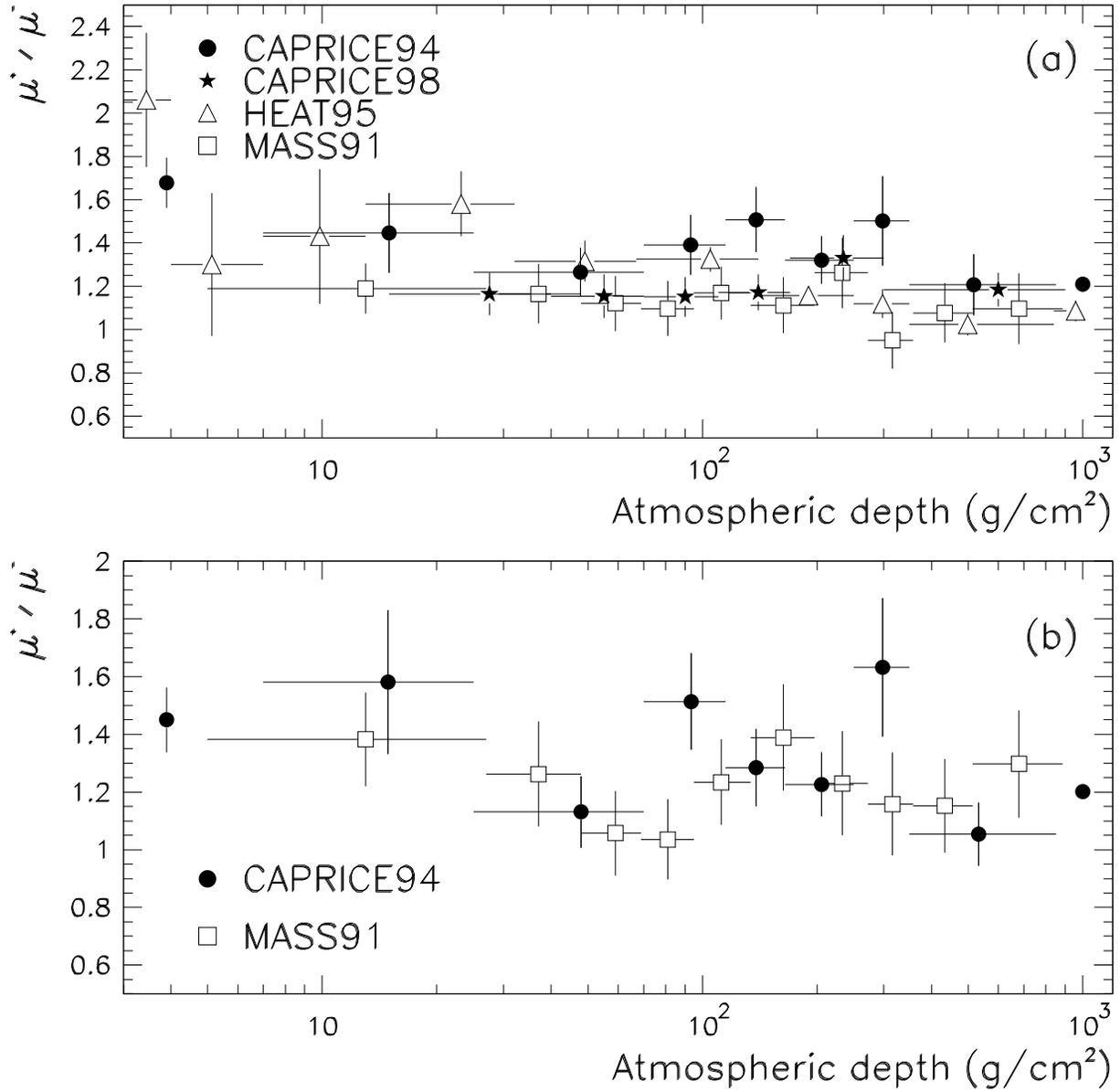,height=18cm}}
\caption{Muon charge ratio as a function of atmospheric depths in two momentum
intervals: (a) 0.3--1~GeV/$c$ in the upper panel and 
1--2~GeV/$c$ in the lower panel (b), obtained in this analysis and from the
CAPRICE98~\protect\cite{cir99}, HEAT95~\protect\cite{cou98} and MASS91
experiments~\protect\cite{bel99}; the latter data refer to the momentum bins
0.3--0.9~GeV/$c$ and 0.9--1.5~GeV/$c$.
\label{muratio} }
\end{figure}

\begin{figure}
\mbox{\epsfig{figure=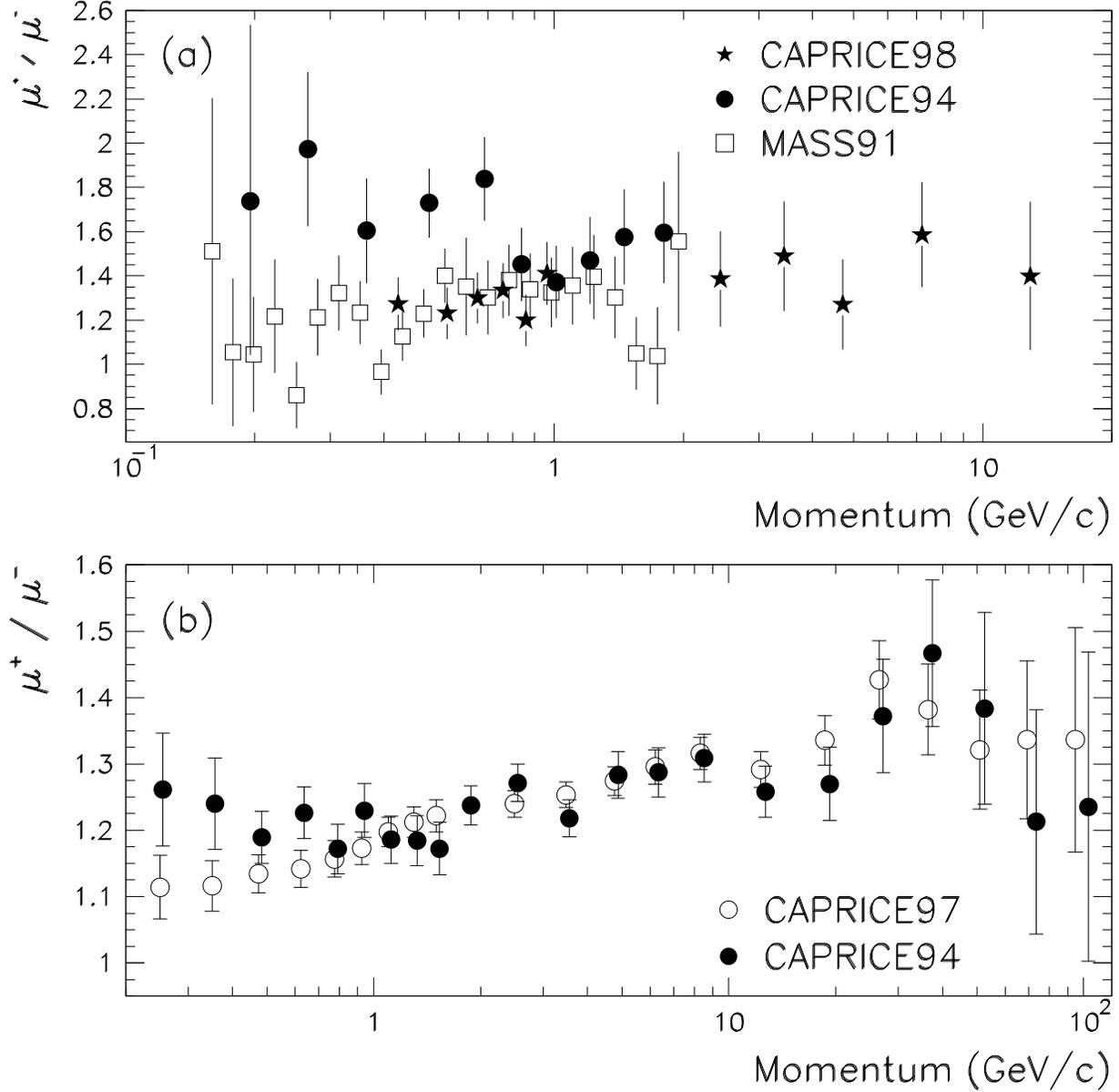,height=18cm}}
\caption{Muon charge ratios as a function of momentum. In the upper panel (a)
are shown the ratios 
obtained in this analysis at 3.9~\gr and those measured
by the 
CAPRICE98 experiment \protect\cite{car99} at 5.5~\gr and with the 
MASS91 experiment \protect\cite{cod97} at 5.7~\gr. In the lower panel (b) are
shown the ratios measured at ground
by this experiment (at 1000~\gr) and by the CAPRICE97 experiment (at 886~\gr) 
\protect\cite{kre99}.
\label{muratio2} }
\end{figure}

Geomagnetic effects also are observed in the muon charge
ratio. Figure~\ref{muratio} shows the \mup to \mum ratio as a function of
atmospheric depth measured in the momentum intervals 0.3--1~GeV/$c$ (a)
and 1--2~GeV/$c$ (b) by this experiment, by the
recent CAPRICE98 experiment \cite{cir99}, which flew from Fort Sumner on 28-29
May 1998, in the range 0.3--0.9~GeV/$c$ by the HEAT95
experiment~\cite{cou98}, which flew from Lynn Lake on the 23rd of August 1995,
and in the range 0.3--0.9~GeV/$c$ (a) and 0.9--1.5~GeV/$c$ (b)
by the MASS91 experiment \cite{bel99}. It can be seen that the CAPRICE94 low
momenta charge ratios are higher than the MASS91 and CAPRICE98 ones. Moreover,
the CAPRICE94 
data show a dependence on the atmospheric depth, which is also visible in the
HEAT data. 
Similar latitude effects also can be seen in Figure~\ref{muratio2}, which shows
(a) the CAPRICE94 data at float along with the charge ratios measured by the
CAPRICE98 experiment \cite{car99} and the MASS91 experiment \cite{cod97} 
and (b) the ground muon data reported here and from the CAPRICE97 experiment
\cite{kre99}, which was carried out in Fort Sumner during Spring 1997.

\begin{figure}
\mbox{\epsfig{figure=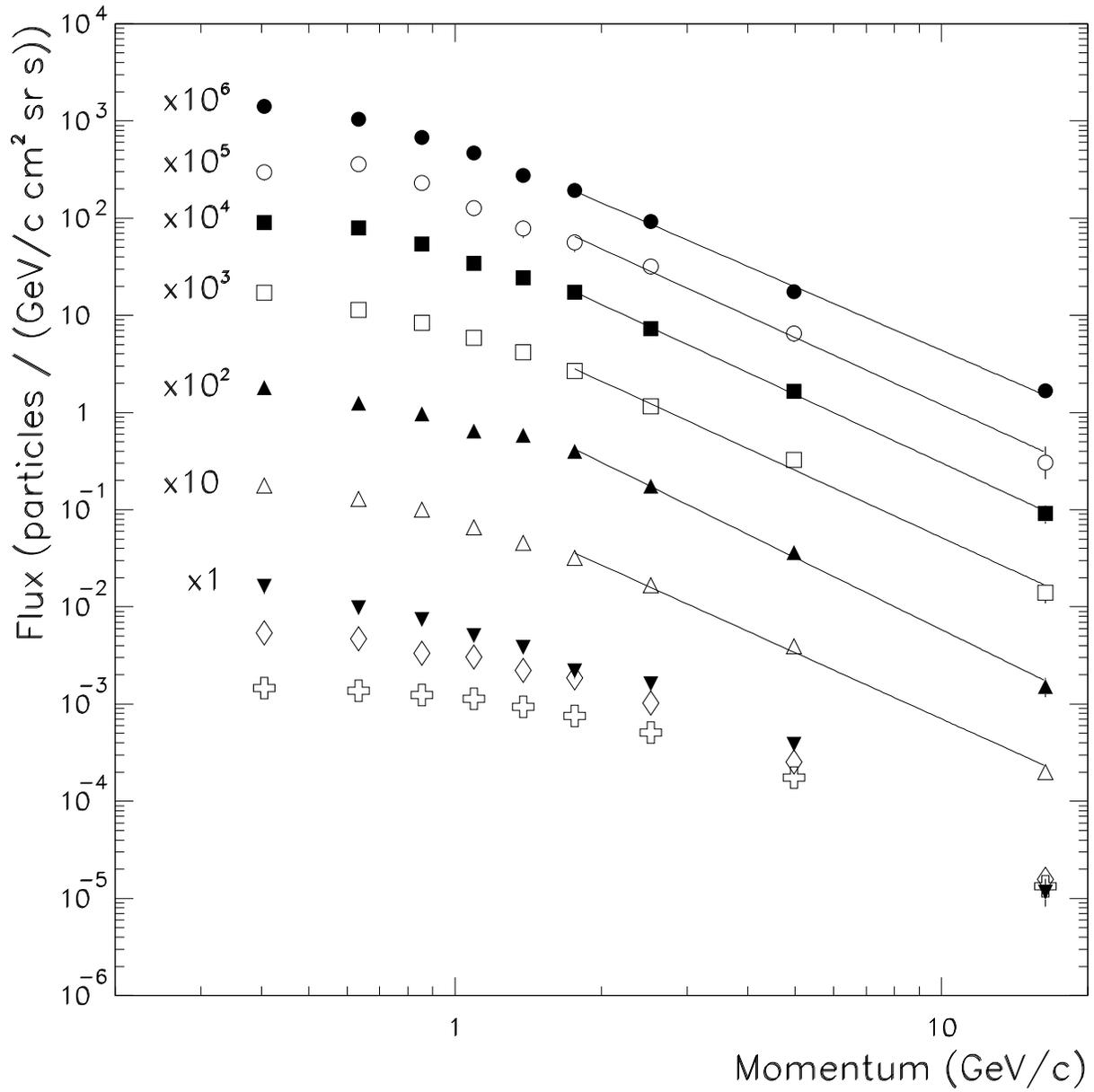,height=18cm}}
\caption{Negative muon spectra for several atmospheric depth intervals.
From top to bottom are the atmospheric depth
ranges in~\gr: 3.3--4.6 (scaled by $10^{6}$), 7--25 (scaled by $10^{5}$),
25--70 ($10^{4}$), 70--115 ($10^{3}$), 115--165 ($10^{2}$),
165--250 (10), 250--350 (1), 350--850 (1) and 1000 (1).
The solid lines are power law fits.
\label{fluxmu} }
\end{figure}

Figure~\ref{fluxmu} shows the measured spectra of negative muons for nine depth
intervals. Above 1.5~GeV/$c$ the \mum spectra between 3.9 and 250~\gr 
(at larger
atmospheric depths unacceptable power law fits were found for this momentum
range) are power law
in momentum with a fairly constant spectral index of $-2.30 \pm 0.04$ that can
be compared with $-2.5 \pm 0.2$ between 20 and 400~\gr~above 
2~GeV/$c$ from MASS89
\cite{bel96} and with $-2.45 \pm 0.05$ between 25 and 250~\gr~above 1.5~GeV/$c$
from MASS91~\cite{bel99}.

\section{Conclusions}
In this paper we have presented new results on atmospheric data measured
with the CAPRICE94 experiment both for positive and negative muons. The data
cover a large atmospheric depth range
from close to the top of the atmosphere (3.9~\gr) down to ground level
(1000~\gr).

The data were compared with other experimental results
\cite{bel99,bel96} that were obtained using the same
superconducting magnet but with different identifying detectors. The muon
spectra 
measured by the different experiments at high momenta (above 1~GeV/$c$) are in
good agreement considering the overall uncertainty of the measurements 
($\sim 10-15 \%$). At lower energy, the comparison between the results of
CAPRICE94 and 
those of MASS89/91 (the CAPRICE94 $\mu^{-}$ fluxes are about 10--20\%
higher than the ones measured in the other
two experiments) indicates solar modulation and geomagnetic
effects. It is worth
pointing out that the differences between the different measurements cannot
account for the discrepancies found at low momenta, while comparing the
experimental data with the theoretical calculation, which are
in some cases as large as 70\% (see \cite{boe99b}).

   \acknowledgements
      This work was supported by NASA Grant NAGW-110, The Istituto Nazionale
      di Fisica Nucleare, Italy, the Agenzia Spaziale Italiana, DARA and
      DFG in Germany, EU SCIENCE, the Swedish National Space Board and
      the Swedish Council for Planning and Coordination of Research.
      The Swedish-French group thanks the EC SCIENCE programme for support.
      We wish to thank the National
      Scientific Balloon Facility and the NSBF launch crew that served
      in Lynn Lake. We would also like to acknowledge the essential support
      given by the CERN TA-1 group and the technical staff of NMSU and of 
      INFN.

\clearpage

\widetext

\begin{table}
\caption{Selection used for identifying muons at different altitudes. See the
text for a detailed discussion about each selection.
\label{t:selmu}}
\begin{tabular}{lccc}
\multicolumn{4}{c}{\mum} \\
\tableline
Selection & Ground & Ascent & Float \\
 & ($X = 1000~\gr$) & ($3.9 < X < 1000~\gr$) & ($X = 3.9~\gr$) \\
\tableline
Tracking & Used & Used & Used \\
dE/dX & Used & Used & Used \\
$\beta_{{\rm tof}}$ & Not used & Not used & Used \\
Calorimeter & Used for $0.3 < R < 3.2$~GV/$c$ &
Used for $R > 0.3$~GV/$c$ & Used for $R > 0.3$~GV/$c$ \\
RICH & All criteria for & All criteria for $R < 0.53$~GV/$c$, & All criteria \\
& $R < 0.53$~GV/$c$ & and criterion 1 for & \\
& & $R > 0.53~$GV/$c, X < 70~\gr$ & \\[0.5cm]
\tableline
\multicolumn{4}{c}{\mup} \\
\tableline
Selection & Ground & Ascent & Float \\
 & ($X = 1000~\gr$) & ($3.9 < X < 1000~\gr$) & ($X = 3.9~\gr$) \\
\tableline
Tracking & Used & Used & Used \\
dE/dX & Used & Used & Used \\
$\beta_{{\rm tof}}$ & Not used & Used & Used \\
Calorimeter & Used for $0.3 < R < 3.2$~GV/$c$ &
Used for $R > 0.3$~GV/$c$  & Used for $R > 0.3$~GV/$c$ \\
RICH & All criteria & All criteria & All criteria \\
\end{tabular}
\end{table}

\clearpage

\begin{table}
\caption{Sources for muon background
\label{t:back}}
\begin{tabular}{lcc}
Source & Rejection criteria & Residual contamination \\
\tableline
Albedo & time-of-flight & none \\
e$^{-}$, e$^{+}$ & RICH and calorimeter & negligible \\
Atmospheric mesons & RICH & negligible for $R < 0.4$~GV/c, \\
& & estimated $< 2\%$ for $R > 0.4$~GV/c \\
Locally produced & RICH, track & at $X < 5~\gr$ \\ 
mesons  & reconstruction & $<$ 20\% for $0.5 < R < 1$~GV/c, \\
& & $<$ 10\% for $R > 1$~GV/c; \\
& & at $X > 7~\gr$ \\
& & negligible for $R < 0.4$~GV/c, \\
 & & $\leq 1\%$ $R > 0.4$~GV/c \\ 
Protons\footnotemark
 & $\beta_{{\rm tof}}$ and RICH; & 
at $X < 5~\gr$ \\ 
& estimated and subtracted & negligible for $R < 1.5$~GV/c, \\  
& at $X < 5~\gr$ for $R > 1.5$~GV/c, & $\leq 1\%$ for $R > 1.5$~GV/c; \\
& at $X = 1000~\gr$ for $R > 3$~GV/c & at $7 < X < 890~\gr$ negligible; \\
& & at $X = 1000~\gr$ \\
& & negligible for $R > 3$~GV/c, \\  
& & $\leq 1\%$ for $R > 3$~GV/c \\ 
\addtocounter{footnote}{-1}
Heavier than p nuclei\footnotemark & dE/dX and RICH & negligible \\
\footnotetext{Only for \mup}
\end{tabular}
\end{table}

\clearpage

\widetext

\begin{table}
\caption{Measured atmospheric growth fluxes (in units of 
particles/(GeV cm$^{2}$ sr s)) for negative and positive muons in
the 0.3 to 40~GeV/$c$ momentum range. Results are given for the following
momentum intervals:
(I) 0.3--0.53~GeV/$c$, (II) 0.53--0.75~GeV/$c$, (III) 0.75--0.97~GeV/$c$, (IV)
0.97--1.23 GeV/$c$, (V) 1.23--1.55 GeV/$c$, (VI) 1.55--2 GeV/$c$, (VII) 2--3.2
GeV/$c$, (VIII) 3.2--8 GeV/$c$, (IX) 8--40 GeV/$c$.
The errors include both statistical and systematic errors.
\label{t:fluxmuon}}
\begin{tabular}{lcccc}
Depth interval & A & B & C & D \\
\tableline
Duration (s) & 85800 & 610 & 620 & 560 \\
$\epsilon_{live}$ & 0.972 & 0.949 & 0.831 & 0.646 \\
Initial depth (\gr) & 1000 & 890 & 580 & 380 \\
Final depth (\gr)  & 1000 & 580 & 380 & 250 \\
\tableline

I  \mum Flux & ($ 1.45 \pm 0.04) \times 10^{-3}$ & ($
  3.49^{+1.09}_{-0.86}) \times 10^{-3}$ & ($
  6.4 \pm 1.3) \times 10^{-3}$ & ($
  1.77 \pm 0.32) \times 10^{-2}$ \\
  FAD (g/cm$^{2}$) &  1000.0 &   709.0 &   463.7 &   308.0 \\
II  \mum Flux & ($ 1.36 \pm 0.03) \times 10^{-3}$ & ($
  2.55 \pm 0.43) \times 10^{-3}$ & ($
  6.67 \pm 0.75) \times 10^{-3}$ & ($
  1.05 \pm 0.16) \times 10^{-2}$ \\
  FAD (g/cm$^{2}$) &  1000.0 &   700.2 &   466.4 &   307.8 \\
III  \mum Flux & ($ 1.23 \pm 0.02) \times 10^{-3}$ & ($
  1.94 \pm 0.37) \times 10^{-3}$ & ($
  4.50 \pm 0.60) \times 10^{-3}$ & ($
  7.26 \pm 1.21) \times 10^{-3}$ \\
  FAD (g/cm$^{2}$) &  1000.0 &   703.2 &   467.0 &   306.2 \\
IV  \mum Flux & ($ 1.13 \pm 0.02) \times 10^{-3}$ & ($
  1.73 \pm 0.32) \times 10^{-3}$ & ($
  4.45 \pm 0.55) \times 10^{-3}$ & ($
  5.35 \pm 0.92) \times 10^{-3}$ \\
  FAD (g/cm$^{2}$) &  1000.0 &   700.7 &   469.0 &   309.1 \\
V  \mum Flux & ($ 9.42 \pm 0.17) \times 10^{-4}$ & ($
  1.34 \pm 0.25) \times 10^{-3}$ & ($
  3.32 \pm 0.42) \times 10^{-3}$ & ($
  3.65 \pm 0.65) \times 10^{-3}$ \\
  FAD (g/cm$^{2}$) &  1000.0 &   701.7 &   470.1 &   310.2 \\
VI  \mum Flux & ($ 7.56 \pm 0.13) \times 10^{-4}$ & ($
  1.21 \pm 0.20) \times 10^{-3}$ & ($
  2.36 \pm 0.30) \times 10^{-3}$ & ($
  2.59 \pm 0.46) \times 10^{-3}$ \\
  FAD (g/cm$^{2}$) &  1000.0 &   707.6 &   471.8 &   309.0 \\
VII  \mum Flux & ($ 5.11 \pm 0.07) \times 10^{-4}$ & ($
  6.63 \pm 0.92) \times 10^{-4}$ & ($
  1.27 \pm 0.14) \times 10^{-3}$ & ($
  1.89 \pm 0.29) \times 10^{-3}$ \\
  FAD (g/cm$^{2}$) &  1000.0 &   708.0 &   469.2 &   310.4 \\
VIII  \mum Flux & ($ 1.75 \pm 0.02) \times 10^{-4}$ & ($
  2.32 \pm 0.28) \times 10^{-4}$ & ($
  2.81 \pm 0.32) \times 10^{-4}$ & ($
  4.34 \pm 0.67) \times 10^{-4}$ \\
  FAD (g/cm$^{2}$) &  1000.0 &   720.0 &   472.3 &   310.2 \\
IX  \mum Flux & ($ 1.33 \pm 0.02) \times 10^{-5}$ & ($
  1.38 \pm 0.26) \times 10^{-5}$ & ($
  1.97 \pm 0.33) \times 10^{-5}$ & ($
  1.37^{+0.43}_{-0.36}) \times 10^{-5}$ \\
  FAD (g/cm$^{2}$) &  1000.0 &   715.6 &   478.1 &   309.1 \\
\tableline

I \mup Flux & ($ 1.76 \pm 0.05) \times 10^{-3}$ & ($
 2.30^{+0.99}_{-0.73}) \times 10^{-3}$ & ($
 8.5 \pm 1.6) \times 10^{-3}$ & ($
 2.28 \pm 0.40) \times 10^{-2}$ \\
 FAD (g/cm$^{2}$) &  1000.0 &   691.9 &   459.2 &   307.3 \\
II \mup Flux & ($ 1.64 \pm 0.03) \times 10^{-3}$ & ($
 2.45^{+0.82}_{-0.64}) \times 10^{-3}$ & ($
 9.6 \pm 1.4) \times 10^{-3}$ & ($
 1.70 \pm 0.29) \times 10^{-2}$ \\
 FAD (g/cm$^{2}$) &  1000.0 &   690.7 &   462.6 &   308.6 \\
III \mup Flux & ($ 1.53 \pm 0.03) \times 10^{-3}$ & ($
 2.08^{+0.73}_{-0.56}) \times 10^{-3}$ & ($
 4.85 \pm 0.92) \times 10^{-3}$ & ($
 1.29 \pm 0.23) \times 10^{-2}$ \\
 FAD (g/cm$^{2}$) &  1000.0 &   703.2 &   462.4 &   307.3 \\
IV \mup Flux & ($ 1.31 \pm 0.02) \times 10^{-3}$ & ($
 1.83^{+0.61}_{-0.47}) \times 10^{-3}$ & ($
 5.49 \pm 0.89) \times 10^{-3}$ & ($
 1.05 \pm 0.19) \times 10^{-2}$ \\
 FAD (g/cm$^{2}$) &  1000.0 &   696.9 &   463.7 &   309.2 \\
V \mup Flux & ($ 1.12 \pm 0.02) \times 10^{-3}$ & ($
 9.7^{+4.2}_{-3.0}) \times 10^{-4}$ & ($
 3.49 \pm 0.63) \times 10^{-3}$ & ($
 5.59 \pm 1.13) \times 10^{-3}$ \\
 FAD (g/cm$^{2}$) &  1000.0 &   692.6 &   464.1 &   309.3 \\
VI \mup Flux & ($ 9.36 \pm 0.16) \times 10^{-4}$ & ($
 1.23^{+0.37}_{-0.29}) \times 10^{-3}$ & ($
 2.00 \pm 0.40) \times 10^{-3}$ & ($
 3.84 \pm 0.78) \times 10^{-3}$ \\
 FAD (g/cm$^{2}$) &  1000.0 &   712.4 &   468.1 &   310.7 \\[0.5cm]

Depth interval & E & F & G & H \\
\tableline
Duration & 740 & 640 & 660 & 690 \\
$\epsilon_{live}$ & 0.516 & 0.449 & 0.396 & 0.358 \\
Initial depth (\gr) & 250 & 190 & 150 & 120 \\
Final depth (\gr)  & 190 & 150 & 120 & 90 \\
\tableline

I  \mum Flux & ($ 1.44 \pm 0.19) \times 10^{-2}$ & ($
  2.06 \pm 0.27) \times 10^{-2}$ & ($
  1.70 \pm 0.25) \times 10^{-2}$ & ($
  1.69 \pm 0.25) \times 10^{-2}$ \\
  FAD (g/cm$^{2}$) &   218.8 &   173.8 &   135.3 &   104.1 \\
II  \mum Flux & ($ 1.26 \pm 0.11) \times 10^{-2}$ & ($
  1.26 \pm 0.13) \times 10^{-2}$ & ($
  1.21 \pm 0.13) \times 10^{-2}$ & ($
  1.21 \pm 0.13) \times 10^{-2}$ \\
  FAD (g/cm$^{2}$) &   219.1 &   174.1 &   135.1 &   104.3 \\
III  \mum Flux & ($ 1.05 \pm 0.10) \times 10^{-2}$ & ($
  8.6 \pm 1.0) \times 10^{-3}$ & ($
  9.1 \pm 1.1) \times 10^{-3}$ & ($
  8.7 \pm 1.1) \times 10^{-3}$ \\
  FAD (g/cm$^{2}$) &   219.3 &   174.3 &   135.0 &   104.1 \\
IV  \mum Flux & ($ 6.51 \pm 0.73) \times 10^{-3}$ & ($
  6.35 \pm 0.83) \times 10^{-3}$ & ($
  6.31 \pm 0.86) \times 10^{-3}$ & ($
  6.05 \pm 0.86) \times 10^{-3}$ \\
  FAD (g/cm$^{2}$) &   219.1 &   174.1 &   135.1 &   104.2 \\
V  \mum Flux & ($ 4.03 \pm 0.52) \times 10^{-3}$ & ($
  6.03 \pm 0.73) \times 10^{-3}$ & ($
  5.15 \pm 0.71) \times 10^{-3}$ & ($
  4.55 \pm 0.67) \times 10^{-3}$ \\
  FAD (g/cm$^{2}$) &   218.3 &   173.7 &   135.5 &   104.4 \\
VI  \mum Flux & ($ 3.36 \pm 0.40) \times 10^{-3}$ & ($
  3.42 \pm 0.46) \times 10^{-3}$ & ($
  3.70 \pm 0.51) \times 10^{-3}$ & ($
  2.92 \pm 0.46) \times 10^{-3}$ \\
  FAD (g/cm$^{2}$) &   219.0 &   173.9 &   135.3 &   104.5 \\
VII  \mum Flux & ($ 1.65 \pm 0.17) \times 10^{-3}$ & ($
  1.49 \pm 0.19) \times 10^{-3}$ & ($
  1.85 \pm 0.22) \times 10^{-3}$ & ($
  1.19 \pm 0.18) \times 10^{-3}$ \\
  FAD (g/cm$^{2}$) &   219.8 &   173.8 &   135.4 &   104.6 \\
\vspace{25cm} \\

VIII  \mum Flux & ($ 3.78 \pm 0.42) \times 10^{-4}$ & ($
  3.66 \pm 0.47) \times 10^{-4}$ & ($
  3.51 \pm 0.49) \times 10^{-4}$ & ($
  3.63 \pm 0.50) \times 10^{-4}$ \\
  FAD (g/cm$^{2}$) &   219.6 &   174.2 &   135.1 &   104.4 \\
IX  \mum Flux & ($ 2.43 \pm 0.41) \times 10^{-5}$ & ($
  1.38^{+0.46}_{-0.35}) \times 10^{-5}$ & ($
  1.42^{+0.49}_{-0.38}) \times 10^{-5}$ & ($
  1.89^{+0.56}_{-0.44}) \times 10^{-5}$ \\
  FAD (g/cm$^{2}$) &   219.8 &   174.9 &   134.6 &   104.6 \\
\tableline

I \mup Flux & ($ 2.07 \pm 0.25) \times 10^{-2}$ & ($
 2.57 \pm 0.32) \times 10^{-2}$ & ($
 2.09 \pm 0.30) \times 10^{-2}$ & ($
 1.85 \pm 0.28) \times 10^{-2}$ \\
 FAD (g/cm$^{2}$) &   219.0 &   174.0 &   135.5 &   104.0 \\
II \mup Flux & ($ 1.68 \pm 0.18) \times 10^{-2}$ & ($
 1.96 \pm 0.23) \times 10^{-2}$ & ($
 1.52 \pm 0.21) \times 10^{-2}$ & ($
 2.06 \pm 0.25) \times 10^{-2}$ \\
 FAD (g/cm$^{2}$) &   219.0 &   174.2 &   135.0 &   104.0 \\
III \mup Flux & ($ 1.17 \pm 0.15) \times 10^{-2}$ & ($
 1.39 \pm 0.19) \times 10^{-2}$ & ($
 1.71 \pm 0.22) \times 10^{-2}$ & ($
 1.24 \pm 0.19) \times 10^{-2}$ \\
 FAD (g/cm$^{2}$) &   219.1 &   173.4 &   135.2 &   104.6 \\
IV \mup Flux & ($ 9.1 \pm 1.2) \times 10^{-3}$ & ($
 9.3 \pm 1.4) \times 10^{-3}$ & ($
 1.05 \pm 0.15) \times 10^{-2}$ & ($
 9.0 \pm 1.5) \times 10^{-3}$ \\
 FAD (g/cm$^{2}$) &   219.5 &   173.8 &   135.1 &   104.4 \\
V \mup Flux & ($ 5.44 \pm 0.83) \times 10^{-3}$ & ($
 7.4 \pm 1.1) \times 10^{-3}$ & ($
 8.0 \pm 1.2) \times 10^{-3}$ & ($
 5.3 \pm 1.0) \times 10^{-3}$ \\
 FAD (g/cm$^{2}$) &   218.7 &   173.4 &   135.5 &   104.3 \\
VI \mup Flux & ($ 2.68 \pm 0.49) \times 10^{-3}$ & ($
 3.56 \pm 0.65) \times 10^{-3}$ & ($
 3.67 \pm 0.69) \times 10^{-3}$ & ($
 4.05 \pm 0.74) \times 10^{-3}$ \\
 FAD (g/cm$^{2}$) &   219.2 &   173.5 &   134.8 &   104.2 \\[0.5cm]

Depth interval & I & J & K & L \\
\tableline
Duration & 720 & 1300 & 1660 & 60520 \\
$\epsilon_{live}$ & 0.327 & 0.303 & 0.289 & 0.269 \\
Initial depth (\gr) & 90 & 65 & 35 & 3.3 \\
Final depth (\gr)  & 65 & 35 & 15 & 4.6 \\
\tableline

I  \mum Flux & ($ 1.66 \pm 0.26) \times 10^{-2}$ & ($
  1.01 \pm 0.14) \times 10^{-2}$ & ($
  4.89 \pm 0.86) \times 10^{-3}$ & ($
  1.40 \pm 0.11) \times 10^{-3}$ \\
  FAD (g/cm$^{2}$) &    77.3 &    50.7 &    25.7 &     3.9 \\
II  \mum Flux & ($ 1.01 \pm 0.12) \times 10^{-2}$ & ($
  8.58 \pm 0.89) \times 10^{-3}$ & ($
  5.56 \pm 0.65) \times 10^{-3}$ & ($
  1.04 \pm 0.06) \times 10^{-3}$ \\
  FAD (g/cm$^{2}$) &    77.2 &    49.8 &    25.1 &     3.9 \\
III  \mum Flux & ($ 8.9 \pm 1.2) \times 10^{-3}$ & ($
  5.81 \pm 0.73) \times 10^{-3}$ & ($
  3.69 \pm 0.52) \times 10^{-3}$ & ($
  6.76 \pm 0.51) \times 10^{-4}$ \\
  FAD (g/cm$^{2}$) &    77.2 &    50.3 &    25.1 &     3.9 \\
IV  \mum Flux & ($ 5.59 \pm 0.84) \times 10^{-3}$ & ($
  3.92 \pm 0.55) \times 10^{-3}$ & ($
  1.73 \pm 0.33) \times 10^{-3}$ & ($
  4.68 \pm 0.38) \times 10^{-4}$ \\
  FAD (g/cm$^{2}$) &    77.2 &    50.6 &    25.9 &     3.9 \\
V  \mum Flux & ($ 3.65 \pm 0.61) \times 10^{-3}$ & ($
  2.64 \pm 0.41) \times 10^{-3}$ & ($
  1.61 \pm 0.29) \times 10^{-3}$ & ($
  2.76 \pm 0.26) \times 10^{-4}$ \\
  FAD (g/cm$^{2}$) &    77.4 &    50.1 &    25.2 &     3.9 \\
VI  \mum Flux & ($ 2.41 \pm 0.42) \times 10^{-3}$ & ($
  2.08 \pm 0.31) \times 10^{-3}$ & ($
  7.6 \pm 1.7) \times 10^{-4}$ & ($
  1.93 \pm 0.19) \times 10^{-4}$ \\
  FAD (g/cm$^{2}$) &    77.2 &    50.4 &    26.2 &     3.9 \\
VII  \mum Flux & ($ 1.17 \pm 0.18) \times 10^{-3}$ & ($
  7.4 \pm 1.1) \times 10^{-4}$ & ($
  4.98 \pm 0.84) \times 10^{-4}$ & ($
  9.19 \pm 0.79) \times 10^{-5}$ \\
  FAD (g/cm$^{2}$) &    77.2 &    50.3 &    25.0 &     3.9 \\
VIII  \mum Flux & ($ 2.61 \pm 0.43) \times 10^{-4}$ & ($
  1.58 \pm 0.26) \times 10^{-4}$ & ($
  1.16 \pm 0.20) \times 10^{-4}$ & ($
  1.76 \pm 0.17) \times 10^{-5}$ \\
  FAD (g/cm$^{2}$) &    77.6 &    50.3 &    24.9 &     3.9 \\
IX  \mum Flux & ($ 7.7 \pm 4.2) \times 10^{-6}$ & ($
  1.06^{+0.34}_{-0.26}) \times 10^{-5}$ & ($
  4.4^{+2.2}_{-1.5}) \times 10^{-6}$ & ($
  1.70 \pm 0.20) \times 10^{-6}$ \\
  FAD (g/cm$^{2}$) &    77.7 &    49.5 &    26.0 &     3.9 \\
\tableline

I \mup Flux & ($ 2.34 \pm 0.33) \times 10^{-2}$ & ($
 1.60 \pm 0.19) \times 10^{-2}$ & ($
 6.4 \pm 1.0) \times 10^{-3}$ & ($
 2.32 \pm 0.18) \times 10^{-3}$ \\
 FAD (g/cm$^{2}$) &    76.9 &    50.7 &    26.1 &     3.9 \\
II \mup Flux & ($ 1.53 \pm 0.22) \times 10^{-2}$ & ($
 1.00 \pm 0.12) \times 10^{-2}$ & ($
 6.63 \pm 0.91) \times 10^{-3}$ & ($
 1.83 \pm 0.11) \times 10^{-3}$ \\
 FAD (g/cm$^{2}$) &    77.5 &    50.2 &    25.1 &     3.9 \\
III \mup Flux & ($ 1.07 \pm 0.18) \times 10^{-2}$ & ($
 5.08 \pm 0.86) \times 10^{-3}$ & ($
 4.05 \pm 0.69) \times 10^{-3}$ & ($
 1.08 \pm 0.08) \times 10^{-3}$ \\
 FAD (g/cm$^{2}$) &    77.6 &    50.6 &    24.7 &     3.9 \\
IV \mup Flux & ($ 7.8 \pm 1.4) \times 10^{-3}$ & ($
 5.10 \pm 0.79) \times 10^{-3}$ & ($
 3.80 \pm 0.62) \times 10^{-3}$ & ($
 5.96 \pm 0.49) \times 10^{-4}$ \\
 FAD (g/cm$^{2}$) &    77.4 &    50.1 &    24.8 &     3.9 \\
V \mup Flux & ($ 6.3 \pm 1.1) \times 10^{-3}$ & ($
 2.50 \pm 0.49) \times 10^{-3}$ & ($
 2.29 \pm 0.43) \times 10^{-3}$ & ($
 4.49 \pm 0.38) \times 10^{-4}$ \\
 FAD (g/cm$^{2}$) &    77.3 &    50.8 &    24.4 &     3.9 \\
VI \mup Flux & ($ 3.15 \pm 0.66) \times 10^{-3}$ & ($
 1.99 \pm 0.37) \times 10^{-3}$ & ($
 1.07^{+0.31}_{-0.24}) \times 10^{-3}$ & ($
 2.91 \pm 0.28) \times 10^{-4}$ \\
 FAD (g/cm$^{2}$) &    77.5 &    50.5 &    25.5 &     3.9 \\

\end{tabular}
\end{table}

\clearpage

\begin{table}
\caption{The $\mu^{-}$ and $\mu^{+}$ fluxes at float altitude (3.9~\gr~of mean
residual atmosphere).
Columns 1 and 2 are, respectively, the rigidity bins in the spectrometer 
and the
mean momenta at the top of the payload. Columns 3 and 4 give the
resulting \mum and \mup fluxes, column 5 the \mup to \mum ratio.
The errors include both statistical and systematic errors.
\label{t:fluxfloat} }
\begin{tabular}{ccccc}
Rigidity & Mean &
\multicolumn{2}{c}{Flux} & \mup / \mum \\
bin & momentum &
\multicolumn{2}{c}{(GeV cm$^{2}$ sr s)$^{-1}$} & \\
GV/$c$ & GeV/$c$ & \mum & \mup & \\
\tableline
  0.15 -  0.20 &       0.19 & (1.66$^{+0.60}_{-0.48}$) $\times 10^{-3}$ &
(2.88 $\pm$ 0.80) $\times 10^{-3}$ & 1.74$^{+0.80}_{-0.70}$ \\
  0.20 -  0.30 &       0.27 & (1.54 $\pm$ 0.20) $\times 10^{-3}$ &
(3.05 $\pm$ 0.35) $\times 10^{-3}$ & 1.97 $\pm$ 0.35 \\
  0.30 -  0.40 &       0.37 & (1.56 $\pm$ 0.18) $\times 10^{-3}$ &
(2.50 $\pm$ 0.23) $\times 10^{-3}$ & 1.60 $\pm$ 0.24 \\
  0.40 -  0.60 &       0.51 & (1.35 $\pm$ 0.10) $\times 10^{-3}$ &
(2.33 $\pm$ 0.12) $\times 10^{-3}$ & 1.73 $\pm$ 0.16 \\
  0.60 -  0.75 &       0.69 & (8.97 $\pm$ 0.76) $\times 10^{-4}$ &
(1.65 $\pm$ 0.10) $\times 10^{-3}$ & 1.84 $\pm$ 0.19 \\
  0.75 -  0.90 &       0.84 & (7.64 $\pm$ 0.70) $\times 10^{-4}$ &
(1.11 $\pm$ 0.08) $\times 10^{-3}$ & 1.45 $\pm$ 0.17 \\
  0.90 -  1.10 &       1.01 & (5.23 $\pm$ 0.49) $\times 10^{-4}$ &
(7.18 $\pm$ 0.52) $\times 10^{-4}$ & 1.37 $\pm$ 0.16 \\
  1.10 -  1.30 &       1.21 & (3.94 $\pm$ 0.42) $\times 10^{-4}$ &
(5.78 $\pm$ 0.47) $\times 10^{-4}$ & 1.47 $\pm$ 0.20 \\
  1.30 -  1.60 &       1.46 & (2.43 $\pm$ 0.27) $\times 10^{-4}$ &
(3.83 $\pm$ 0.31) $\times 10^{-4}$ & 1.58 $\pm$ 0.21 \\
  1.60 -  2.0 &       1.80 & (1.73 $\pm$ 0.19) $\times 10^{-4}$ &
(2.75 $\pm$ 0.25) $\times 10^{-4}$ & 1.60 $\pm$ 0.23 \\
  2.0 -  3.0 &       2.46 & (9.87 $\pm$ 0.96) $\times 10^{-5}$ & & \\
  3.0 -  5.0 &       3.87 & (3.37 $\pm$ 0.39) $\times 10^{-5}$ & & \\
  5.0 - 10.0 &       6.97 & (8.60 $\pm$ 1.22) $\times 10^{-6}$ & & \\
 10.0 - 20.0 &      13.90 & (2.07 $\pm$ 0.42) $\times 10^{-6}$ & & \\

\end{tabular}
\end{table}

\clearpage

\clearpage

\clearpage

\clearpage

\clearpage

\clearpage

\clearpage

\clearpage

\clearpage

\clearpage

\clearpage

\clearpage

\clearpage

\end{document}